\documentclass[useAMS,usegraphicx]{mn2e}
\usepackage{pdflscape}
\usepackage{epsfig}
\usepackage[version=3]{mhchem}
\usepackage{tablefootnote}
\usepackage{url}
\usepackage{amssymb}
\usepackage{amsmath}
\usepackage{multicol}
\newcommand{\src} {IRAS\,20126+4104}
\newcommand{\Tdust}    {T_\mathrm{dust}}
\newcommand{\Tex}   {T_\mathrm{ex}}
\newcommand{\Trot}  {T_\mathrm{rot}}
\newcommand{\mum}   {$\mu$m}
\newcommand{\kms}   {km~s$^{-1}$}

\newcommand{\cmt}   {cm$^{-3}$}
\newcommand{\jpb}   {$\rm Jy~beam^{-1}$}    
\newcommand{\lo}    {$L_{\sun}$}
\newcommand{\mo}    {$M_{\sun}$}

\newcommand{\tco}    {$^{13}$CO}

\newcommand{\methanol}  {CH$_3$OH}

\newcommand{\formald} {H$_2^{\phantom,13}$CO}
\newcommand{\ethyleneg}  {(CH$_2$OH)$_2$}

\newcommand{\acetone} {CH$_3$COCH$_3$}
\newcommand{\dimethylether} {CH$_3$OCH$_3$}
\newcommand{\acetaldehyde} {CH$_3$CHO}
\newcommand{\methylformate} {HCOOCH$_3$}

\newcommand{\methylcyanide} {CH$_3$CN}

\newcommand{\ethylcyanide} {CH$_3$CH$_2$CN}
\newcommand{\et}    {et al.}
\newcommand{\eg}    {e.\,g.,}
\newcommand{\ie}    {i.\,e.,}

\newcommand{\supa}  {$^\mathrm{a}$}
\newcommand{\supb}  {$^\mathrm{b}$}
\newcommand{\supc}  {$^\mathrm{c}$}
\newcommand{\supd}  {$^\mathrm{d}$}
\newcommand{\supe}  {$^\mathrm{e}$}
\newcommand{\supf}  {$^\mathrm{f}$}
\newcommand{\supg}  {$^\mathrm{g}$}
\newcommand{\suph}  {$^\mathrm{h}$}
\newcommand{\supi}  {$^\mathrm{i}$}

\newcommand{\phe}   {\phantom{$^\mathrm{c}$}}

\title[COMs in IRAS\,20126+4104 at $\sim600$~au]{Complex Organic Molecules tracing shocks along the outflow cavity in the high-mass protostar IRAS\,20126+4104}
\author[Palau et al.]{Aina Palau$^{1}$\thanks{E-mail:a.palau@crya.unam.mx}, 
Catherine Walsh$^{2,3}$, \'Alvaro S\'anchez-Monge$^{4}$, Josep M. Girart$^{5}$, 
 \newauthor
Riccardo Cesaroni$^{6}$, 
Izaskun Jim\'enez-Serra$^{7}$,
Asunci\'on Fuente$^{8}$, 
Luis A. Zapata$^{1}$,
\newauthor
Roberto Neri$^{9}$
\\
\\
$^{1}$ Instituto de Radioastronom\'ia y Astrof\'isica, Universidad Nacional Aut\'onoma de M\'exico, P.O. Box 3-72, 58090, Morelia, Michoac\'an, M\'exico\\
$^{2}$ Leiden Observatory, Leiden University, P.O. Box 9513, NL-2300 RA Leiden, The Netherlands\\
$^{3}$ School of Physics and Astronomy, University of Leeds, Leeds UK, LS2 9JT\\
$^{4}$ I. Physikalisches Institut der Universit\"at zu K\"oln, Z\"ulpicher Strasse 77, 50937 K\"oln, Germany\\
$^{5}$ Institut de Ci\`encies de l'Espai (CSIC/IEEC), Campus UAB, Facultat de Ci\`encies, Torre C-5 parell 2, E-08193 Bellaterra, Catalunya, Spain\\  
$^{6}$ Osservatorio Astrofisico di Arcetri, INAF, Lago E. Fermi 5, 50125, Firenze, Italy\\
$^{7}$ School of Physics and Astronomy, Queen Mary University of London, Mile End Road, London E1 4NS, UK\\
$^{8}$ Observatorio Astron\'omico Nacional, P.O. Box 112, 28803 Alcal\'a de Henares, Madrid, Spain\\
$^{9}$ Institut de Radioastronomie Millim\'etrique, Grenoble, France\\
}
\begin{document}

\date{Accepted date. Received date; in original form date}

\pagerange{\pageref{firstpage}--\pageref{lastpage}} \pubyear{2016}

\maketitle

\label{firstpage}

\begin{abstract}
We report on subarcsecond observations of complex organic molecules (COMs) in the high-mass protostar \src\ with the Plateau de Bure Interferometer in its most extended configurations. In addition to the simple molecules SO, HNCO and \formald, we detect emission from \methylcyanide, \methanol, HCOOH, \methylformate, \dimethylether, \ethylcyanide, \acetone, NH$_2$CN, and \ethyleneg. 
SO and HNCO present a X-shaped morphology consistent with tracing the outflow cavity walls. 
Most of the COMs have their peak emission at the putative position of the protostar, but also show an extension towards the south(east), coinciding with an H$_2$ knot from the jet at about 800--1000~au from the protostar.
This is especially clear in the case of \formald\ and \dimethylether.
We fitted the spectra at representative positions for the disc and the outflow, and found that the abundances of most COMs are comparable at both positions, suggesting that COMs are enhanced in shocks as a result of the passage of the outflow.
By coupling a parametric shock model to a large gas-grain chemical network including COMs, we find that the observed COMs should survive in the gas phase for $\sim2000$~yr, comparable to the shock lifetime estimated from the water masers at the outflow position.
Overall, our data indicate that COMs in \src\  may arise not only from the disc, but also from dense and hot regions associated with the outflow.
\end{abstract}


\section{Introduction}\label{intro}

A common signature associated with the earliest stages of high-mass star formation is the presence of hot molecular cores, which are compact ($\le\!0.05$~pc) objects with high temperatures ($\ga100$~K) and densities ($n\ga\!10^6$~cm$^{-3}$), characterized by a very rich chemistry of complex organic molecules (COMs, molecules with $>6$ atoms, \eg\ Blake et al. 1987, Kurtz 2005, Herbst \& van Dishoeck 2009, Ob\"erg et al. 2014). 
Such a rich chemistry is supposed to be triggered first during a cold and dense phase which favors the formation of simple hydrogenated species on the surface of dust grains. In a subsequent `warm-up' phase, radicals are efficiently recombined on the grain surfaces (\eg\ Garrod \& Herbst 2006; Garrod et al. 2008; Aikawa et al. 2008), and finally the heating from the nascent massive star evaporates the complex molecules from the dust grain mantles and triggers additional gas-phase reactions (\eg\  Millar \et\ 1997). However, these models fail to predict by more than one order of magnitude the abundances of well-measured COMs, such as  \methylformate\ and \dimethylether\  (\eg\ Taquet et al. 2012, Lykke et al. 2016), to the point that very recent works are starting to propose that the role of gas-phase chemistry might be more important than previously thought (\eg\ Balucani, Ceccarelli, \& Taquet 2015, Enrique-Romero et al. 2016).

In addition, COMs have unambiguously been detected in cavities of outflows powered by low-mass ($\la2$~\mo) protostars (\eg\ L1157-B1: Arce et al. 2008; Codella et al. 2015; IRAS\,16293$-$2422: Chandler et al. 2005), and  intermediate-mass ($2$--8~\mo) protostars (\eg\ Orion-KL: Liu \et\ 2002; Favre \et\ 2011a; Zapata et al. 2011; NGC\,2071: van Kempen et al. 2014). 
Actually, chemical modelling indicates that photodissociation in conjunction with reactive desorption in the outflow cavity walls could be a relevant process to enhance gas phase abundances of COMs (Drozdovskaya et al. 2015), although this has been tested only in the low-mass case.

Recently, several works have studied the morphology of the COM emission at sufficiently high spatial resolution to disentangle the possible origin from the disc and/or outflow. In particular, Palau et al. (2011) present emission from COMs towards three intermediate-mass hot cores (one in IRAS\,22198+6336, two in AFGL\,5142) down to disc spatial scales ($\sim500$~au), and find that in two cases the COM emission is resolved in the disc direction, suggesting that COMs are tracing protostellar discs in these cases. 
On the other hand, Fuente et al. (2014) did not resolve the emission from several COMs down to a spatial resolution of 1700~au towards NGC\,7129-FIRS2, leaving open the possibility that some COMs are tracing shocks in the outflow.
However, these hot cores are associated with intermediate-mass protostars, and thus their UV radiation and/or outflow momentum might be too faint to efficiently heat and evaporate the molecules in the outflow cavity walls. 

We present here subarcsecond interferometric observations of COMs towards the prototypical high-mass protostar IRAS\,20126+4104, with a luminosity of $\sim8900$~\lo, a mass of $\sim12$~\mo\ (Chen et al. 2016), and located at a distance of 1.64~kpc (\eg\ Cesaroni \et\ 1999; Moscadelli \et\ 2011). This allowed us to resolve the emission from COMs, of up to 10 atoms, down to $\sim600$~AU for this high-mass protostar, adding crucial information about the possible origin of COM emission. This is important, because the object is known to drive a highly-collimated bipolar outflow (opening angle $\sim9^\circ$), oriented in the southeast-northwest direction (with position angle P.A.$=115^\circ$), and being almost on the plane of the sky (Moscadelli et al. 2011). In addition, the jet structure associated with this outflow has been recently studied at subarcsecond angular resolution, revealing that the jet is asymmetric, with the southeastern lobe being probably older than the northwestern one, and being already associated with a cavity bright in H$_2$ at 2.12~\mum, located at about 800--1000~au from the protostar (Cesaroni et al. 2013). On the other hand, the northwestern lobe should be still at the very beginning of the creation of the cavity. Therefore, the subarcsecond observations presented here constitute an excellent dataset to study the emission of COMs at an angular resolution that allows to disentangle the emission coming from the jet  and the emission coming from the disc. In Section~2, we describe the observations. In Section~3, we present the observed spectra towards \src\ with the line identification and integrated emission images of the strongest detected COMs. In Section~4, we study the morphology of the COM emission relative to the dust continuum, and present a shock model coupled to a gas-grain chemical network used to explain our findings. In Section~5 we discuss the results, and in Section~6 we present our main conclusions.

\section{Observations}\label{obs}

The dataset presented in this work is part of the dataset reported in Cesaroni et al. (2014), which is focused on the 1.4~mm continuum and CH$_3$CN\,(12--11) emission. We provide here only a brief description of the observations, and refer the reader to Cesaroni et al. (2014) for further details.
The IRAM Plateau de Bure Interferometer (PdBI\footnote{IRAM is supported by INSU/CNRS (France), MPG (Germany) and IGN (Spain).}) was used in A and B configurations to observe the continuum and line emission at 220.500~GHz of \src. 
Observations were carried out on 2008 January 29, February 13 and 16, and March 15, with the phase center being $\alpha$(J2000)=20:14:26.036,   $\delta$(J2000)=41:13:32.52.
Phase and bandpass calibration were performed using 2013+370 and 3C273, yielding a phase rms noise of $30^\circ$.
This corresponds to an uncertainty in absolute position of $0.03''$--$0.05''$.
The absolute flux density scale was determined from MWC349.

Two correlator units of 160~MHz of bandwidth with 256 spectral channels were used to observe the \methylcyanide\,(12--11) lines in each polarization (Cesaroni et al. 2014). 
Four additional units (per polarization) of 320~MHz bandwidth with 128 channels were used to observe the continuum across $\sim1.2$~GHz. This, together with the aforementioned units of 160 MHz, allowed us to cover a total frequency range of 1.65~GHz.
Most of the COMs were detected in the 320~MHz units, which provide a spectral resolution of 2.5~MHz or ~3.4~\kms. 

Calibration and imaging were performed using the \textsc{GILDAS} software package\footnote{GILDAS: Grenoble Image and Line Data
Analysis System, see http://www.iram.fr/IRAMFR/GILDAS.}, using natural weighting over the entire frequency range.
The  synthesized beam is $0.46''\times0.32''$ at P.A.=$56^\circ$, and the rms of the final cleaned maps is $\sim3$~m\jpb\ per channel. 
The conversion factor from \jpb\ to K is 170.5~K\,Jy$^{-1}$\,beam.

\begin{figure*}
\begin{center}
\begin{tabular}[b]{c}
    \epsfig{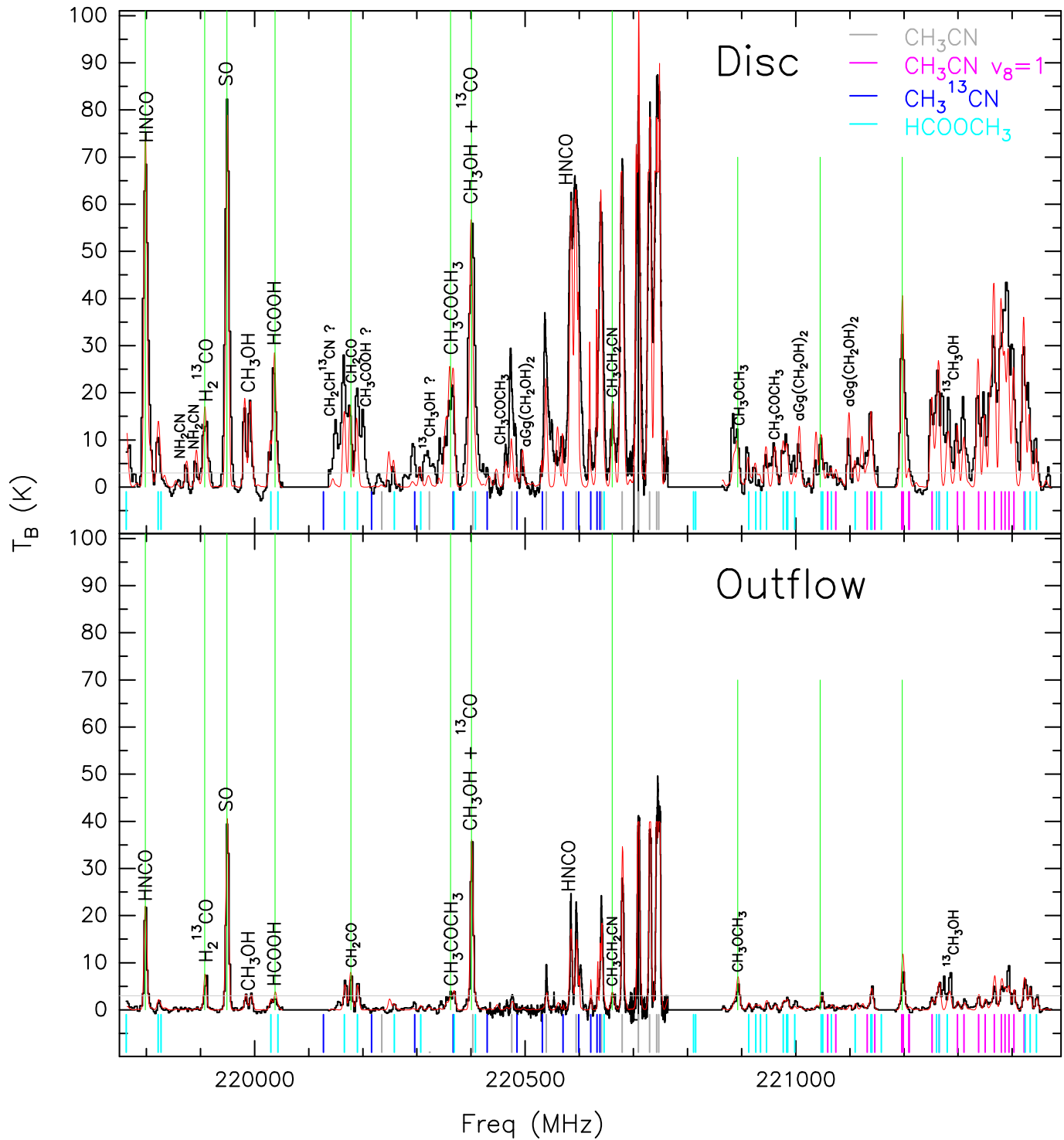}\\
\end{tabular}
\caption{{\bf Top:} PdBI beam averaged spectrum towards the peak of the continuum of \src\ (the putative position of the disc, marked in Fig.~\ref{fm0simple}a with green dots).
{\bf Bottom:} idem towards the `outflow' position (see main text, marked in Fig.~\ref{fm0simple}a with green dots).
The red line corresponds to the synthetic spectrum resulting from summing the individual synthetic spectra fitted using XCLASS. Vertical green lines mark the transitions whose zeroth-order moment has been imaged and is shown in Figs.~\ref{fm0simple} and \ref{fm0coms}.
The grey horizontal line marks the 5$\sigma$ detection threshold. Labels indicate the molecule which dominates the emission in each spectral line.
}
\label{fspec}
\end{center}
\end{figure*}

\section{Results}\label{res}

\begin{figure}
\begin{center}
\begin{tabular}[b]{c}
    \epsfig{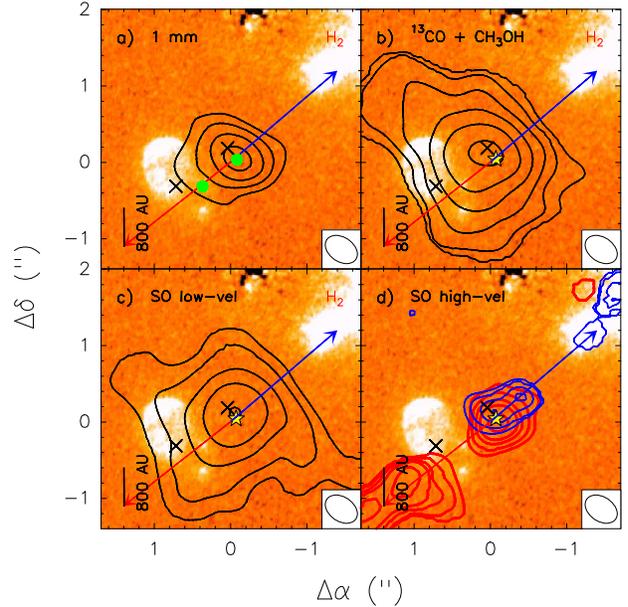}\\
\end{tabular}
\caption{a) PdBI 1.4~mm continuum image (Cesaroni et al. 2014); b) zeroth-order moment for \tco\ + \methanol; c) zeroth-order moment for SO, integrated from $-8$ to +8~\kms\ with respect to the systemic velocity (low-velocity); d) zeroth-order moment for SO, integrated from 11 to 32~\kms\ with respect to the systemic velocity (high-velocity, red contours), and from $-26$ to $-12$~\kms\ with respect to the systemic velocity (high-velocity, blue contours). 
The transitions corresponding to panels `b', `c', and `d' are marked with green vertical lines in Fig.~1, and their frequency and quantum numbers are given in Table~\ref{tcoms}. 
In all panels, the two crosses correspond to the $K$-band infrared sources reported by Sridharan \et\ (2005) tracing the outflow cavities, the color image corresponds to the emission of H$_2$ at 2.12~\mum\ from Cesaroni \et\ (2013), also tracing the (south-eastern) outflow cavity (Cesaroni \et\ 2013), and the star symbol marks the position of the 1.4~mm peak (panel `a').
The two green dots in panel `a' indicate the positions (peak, outflow) where the spectra shown in Fig.~1 have been extracted. 
\vspace{2cm}
}
\label{fm0simple}
\end{center}
\end{figure}

Figure~\ref{fspec} presents the beam averaged spectrum in the frequency range 219750--221400~MHz for the two positions indicated in Fig.~\ref{fm0simple}a with green dots. 
One position is centred on the peak of the continuum emission (the putative position of the disc, Cesaroni et al. 2014) and the other position is centred along the outflow direction, about $\sim1$~arcsec to the southeast.
In order to identify the lines detected above $5\sigma$ (see Fig.~\ref{fspec}), we used the XCLASS software (M\"oller et al.\ 2015) to generate synthetic spectra that are fitted to the observed spectra. XCLASS assumes Local Thermodynamic Equilibrium\footnote{XCLASS: eXtended CASA Line Analysis Software Suite, considers that one excitation temperature describes the different line transitions for each model component.}, 
and takes into account opacity effects and line blending from multiple species and components. XCLASS uses the spectroscopic data saved in the VAMDC\footnote{Virtual Atomic and Molecular Data Centre, \url{http://vamdc.eu/}} portal, combining the CDMS (Cologne Database for Molecular Spectroscopy; M\"uller et al.\ 2005) and JPL (Jet Propulsion Laboratory; Pickett et al.\ 1998) spectral line catalogues. The main adjustable parameters in the creation of synthetic spectra are the source size, the rotational temperature, the column density of the molecule, the line width and the velocity of the source. We used the model optimizer MAGIX (M\"oller et al.\ 2013) within XCLASS to fit the observed spectra with the synthetic spectra generated including the molecules responsible for each detected line (listed in Table~\ref{tcoms}). To obtain reasonable fits for \methylcyanide, we had to use two components: a foreground cold component and an inner hotter component.

For the adjustable parameters, we assumed a filling factor equal to unity, and a linewidth of 8.0~\kms\ (for all the components, except for $^{13}$CO, for which we used 15.0~ \kms, and for the foreground component of CH$_3$CN for which we used 2.5~\kms). The line width of 8.0~\kms\ was obtained from the average of line widths measured for the strongest detected lines. The adopted $V_\mathrm{LSR}$ for each molecule are listed in Table~\ref{tcoms}, and are obtained from a Gaussian fit to the strongest line of each molecule, in the case the molecule is not blended and is unambiguously identified, or from a simultaneous fit to all the lines of \methylformate\ for the other cases, as \methylformate\ is one of the molecules with highest number of detected transitions in our observed frequency range. The column density ratios CH$_3$OH/$^{13}$CH$_3$OH and CH$_3$CN/CH$_3^{13}$CN were allowed to vary from 30 to 60 (\eg\ Wilson \& Rood 1994; Visser et al. 2009; J{\o}rgensen et al. 2016; Li et al. 2016), and the best fits correspond to values of these ratio abundances around 30.
The rotational temperature was left as a free parameter (ranging from 50 to 300~K, Cesaroni et al.\ 1997, Isokoski et al.\ 2013) for CH$_3$CN, \acetone, \methylformate, and NH$_2$CN (only for the disc) as these are the molecules with highest number of transitions detected in the observed frequency range. 

The averaged value of the fitted rotational temperatures, for the outflow and peak positions, is used as an input for the remaining molecules\footnote{For $^{13}$CO, likely tracing more extended material, we assumed a temperature of 100~K.}, leaving the column density as the only free parameter in those cases.
Table~\ref{tcoms} lists the temperatures and column densities used to generate the synthetic spectra. 
The errors of the fitted parameters calculated by XCLASS are in the range 15--30~K for the temperature, and 0.1--1 for the logarithm of the column density (in cm$^{-2}$).

Figure~\ref{fspec} shows the total fitted synthetic spectra (summing up the contributions of all the identified lines), and in Appendix A we provide individual synthetic spectra for each molecular species\footnote{The total synthetic spectrum of Fig.~\ref{fspec} for the disc position presents a clear deficit of emission at some frequencies with respect to the observed spectrum, especially at around 220150 and 220199~MHz. We searched for possible transitions responsible for this excess of observed emission and found that for 220150~MHz this excess could be due to CH$_2$CH$^{13}$CN (vinyl cianyde). However, since we did not find evidences of the corresponding main isotope in our spectrum, this identification remains tentative. As for the emission at 220199~MHz, this could be due to CH$_3$COOH (acetic acid). In this case, the molecule is not included in the CDMS database and prevents us from generating the synthetic spectrum with XCLASS, remaining its identification tentative as well.}.

\begin{figure*}
\begin{center}
\begin{tabular}[b]{c}
    \epsfig{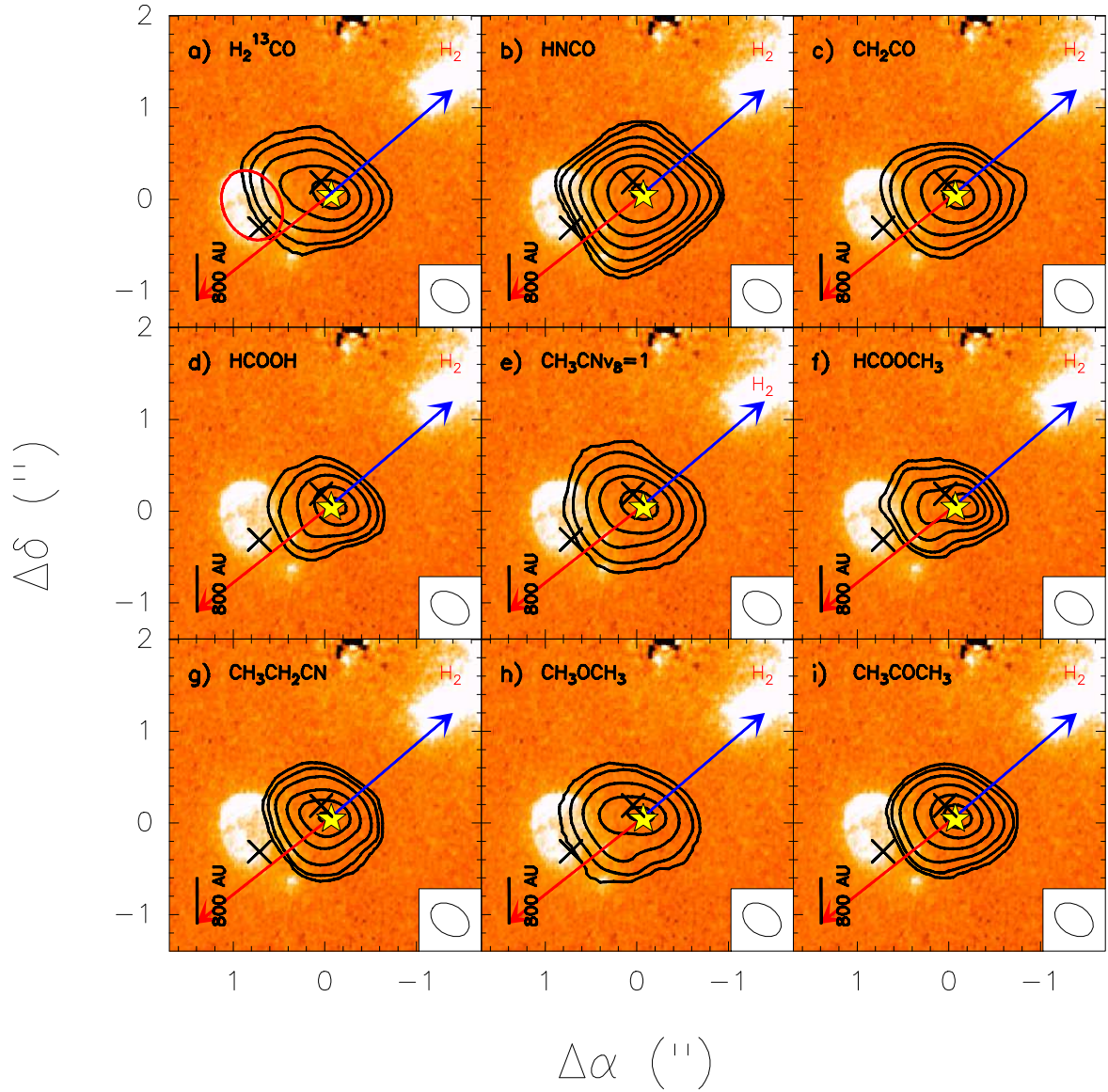}\\
\end{tabular}
\caption{PdBI zeroth-order moment maps for a number of transitions (marked with green vertical lines in Fig.~\ref{fspec}, and given in Table~\ref{tcoms}) from COMs detected towards \src. 
In all panels, the two crosses correspond to the $K$-band infrared sources reported by Sridharan \et\ (2005) tracing the outflow cavities, and the color image corresponds to the emission of H$_2$ at 2.12~\mum\ from Cesaroni \et\ (2013), also tracing the (south-eastern) outflow cavity (Cesaroni \et\ 2013, marked with a red ellipse in panel `a' for further reference).
The star symbol marks the position of the 1.4~mm peak from Cesaroni et al. (2014, also shown in Fig.~\ref{fm0simple}a).
\vspace{2cm}
}
\label{fm0coms}
\end{center}
\end{figure*}

Figures~\ref{fm0simple} and \ref{fm0coms} present the zeroth-order moment of the strongest detected molecular lines.
The simplest molecules are \tco\ and SO.
The SO emission shows two velocity components. One is a low-velocity component that traces the same velocity range as the other molecular lines (integrated from  $-8$ to +8~\kms\ with respect to the systemic velocity, $V_\mathrm{LSR}\sim-3.5$~\kms, Moscadelli et al. 2011; Isokoski et al. 2013). The other one is a high velocity component (integrated up to $\sim\pm30$~\kms\ with respect to the systemic velocity). 
While the low-velocity SO emission presents an extended X-shaped morphology (see Fig.~\ref{fm0simple}c), the high-velocity emission traces the well-known collimated outflow driven by \src, with the redshifted lobe towards the southeast and the blueshifted lobe towards the northwest of the continuum source. The P.A. of the SO outflow is $\sim130$\degr, similar to the P.A. obtained by Moscadelli et al. (2011) from water masers, of $\sim115$\degr\ (Fig.~\ref{fm0simple}d). Both high-velocity redshifted and blueshifted emission overlap at the position of the millimeter source. 

Slightly more complex molecules (with up to 4 atoms) are H$_2$$^{13}$CO and HNCO.
The HNCO emission also presents a X-shaped structure, similar to SO, although more compact (Fig.~\ref{fm0coms}b). On the other hand, H$_2$$^{13}$CO is better resolved in the southwest-northeast direction, \ie\ perpendicular to the outflow (Fig.~\ref{fm0coms}a, P.A.$\sim75$\degr).

Concerning even more complex molecules, we detected two molecules with 5 atoms (HCOOH, CH$_2$CO), four molecules with 6 atoms (\methanol, \methylcyanide, and both $^{13}$C isotopologues), 
and four molecules containing between 8 and 10 atoms (\methylformate, \ethylcyanide, \dimethylether, and \acetone). We would like to emphasise that all these molecules are detected \emph{both} at the disc and at the outflow positions (Fig.~\ref{fspec}).
At the position of the disc we additionally detected NH$_2$CN, and \ethyleneg.
The molecule with the highest number of strong transitions in the covered frequency range is \methylformate\ (see Fig.~A11 of Appendix A). 
The integrated emission of these molecules consists of a centrally-peaked structure at the position of the millimeter continuum source (Fig.~\ref{fm0coms}) with an elongation towards the south(east), especially clear in \methylformate, \dimethylether, and HCOOH. An elongation towards the south is also notable in \methylcyanide\ (Cesaroni et al. 2014), and is attributed to an inhomogeneous medium surrounding \src, which should be less dense to the south because the outflow seems to have already created a cavity (Cesaroni et al. 2013, 2014).

Table~\ref{tsizes} lists the results of a 2D-Gaussian fit to the integrated emission (using the task `JMFIT' from the AIPS package) of the strongest lines. 
The measured sizes range from 600 to 1200~AU, with P.A.$\ga90$\degr\ for all the cases, except for H$_2$$^{13}$CO. In particular, the P.A. of HCOOH, \methylformate, \dimethylether, and \acetone\ are 100--110\degr. This orientation is close to the P.A. measured for the jet from water masers ($\sim115$\degr, Moscadelli et al. 2011). Although the uncertainties in the 2D-Gaussian fits are in some cases large (0.1--0.2 arcsec), the residual maps still indicate a clear excess of emission to the south(east).

In Table~\ref{tcoms} we list the fitted/adopted rotational temperature $\Trot$ and the fitted column density $N$ obtained by XCLASS for both the disc and outflow positions. 
$\Trot$ ranges from 210 to 250~K for the disc position, and from 170 to 190~K for the outflow position (excluding the fit to CH$_3$CN), consistent with Cesaroni et al. (1997) and Chen et al. (2016). 
To estimate the average $\Trot$ we did not take into account the fit to CH$_3$CN because we used two components in this case. The resulting average $\Trot$ is 230~K for the disc, and 180~K for the outflow. 
The column densities at the disc position are around $10^{16}$--$10^{17}$~cm$^{-2}$, while those at the outflow position are about one order of magnitude lower. 

In order to test the LTE assumption, especially at the outflow position where the density is not as high as in the disc, we used the radiative transfer code RADEX (van der Tak et al. 2007) with the temperature and column density inferred for the outflow position (180 K, and Table 1) and a linewidth of 8~\kms, for the transitions of the molecules detected by us (Table 1) and available in this code (HNCO, H$_2$CO, CH$_3$CN)\footnote{For CH$_3$OH we refrained from using RADEX because our specific combination of parameters yielded instabilities in the code. In addition, the critical density of \methanol\ is only $\sim10^4$~\cmt\ (see Table~1), and hence it should be well thermalized for the range of densities inferred from the other molecules ($10^4$--$10^5$~\cmt).}. For these transitions, we determined the densities required to reach brightness temperatures comparable to those observed, and found that densities in the range $10^4$--$10^5$~\cmt\ yield brightness temperatures in the range 17--92~K, consistent with our observations\footnote{For the case of a density of $10^4$~\cmt, we need in general column densities in RADEX which are a factor of a few larger than those derived using XCLASS. Thus, in the case that the true density at the outflow position was $\sim10^4$~\cmt, the LTE assumption would yield lower limit column densities by a factor of a few, in full agreement with the work of Faure et al. (2014).}. 
Since the densities required to detect those transitions are easily reached at such close distances of a high-mass protostar (see, \eg\ Palau et al. 2014), the LTE assumption seems to be appropriate for both the disc and the outflow positions. This was also hinted by the fact that the inferred rotational temperatures at both positions are high, of the order of $\sim200$~K.

Finally, we estimated the column density of H$_2$ by using the 1.4~mm continuum emission shown in Fig.~\ref{fm0simple}a (Cesaroni \et\ 2014), for the disc and outflow positions. To do this, we integrated the 1.4~mm continuum emission inside the same polygon where the spectra were extracted, and found 51~mJy for the disc and 5.7~mJy for the outflow. To estimate the corresponding mass of gas and dust, we assumed optically thin emission, a dust temperature equal to the average rotational temperature inferred by XCLASS for the disc and outflow positions (see previous paragraph), and a dust opacity at 1.4~mm of 0.7981~cm$^2$\,g$^{-1}$ (inferred from the opacity tables of Ossenkopf \& Henning 1994, for the case of thin ice mantles at a density of 10$^6$~\cmt). 
Dividing the derived mass by the area of the polygon with which we measured the 1.4~mm flux density, we obtained an H$_2$ column density of $2.7\times10^{24}$~cm$^{-2}$ for the disc, and $5.3\times10^{23}$~cm$^{-2}$ for the outflow position.
With these H$_2$ column densities, we calculated the abundances for each COM for the disc and the outflow, finding values typically in the range $10^{-9}$--10$^{-8}$ (see Table~\ref{tcoms}), fully consistent with the CH$_3$CN and CH$_3$OH abundances derived by Chen et al. (2016) at similar angular resolution for this source.
It is interesting to note that in most cases the abundance of the COM is very similar or even higher at the outflow position than at the disc position, with the clearest case being \dimethylether, for which the abundance in the outflow is a factor of 4 higher than the abundance in the disc. The only COM presenting an abundance clearly higher (about 3 times) in the disc than in the outflow is \acetone, suggesting an anticorrelation between \acetone\ and \dimethylether.


\begin{landscape}
\begin{table}
\caption{Properties of the emission of complex organic molecules from IRAS\,20126+4104 observed with a spatial resolution of $\sim600$~au, for two positions representative of the disc and the outflow. Molecules are sorted by increasing number of atoms (increasing complexity).}
\centering
\begin{tabular}{l l c c c c c c c c c c c c c c}
\hline\hline\noalign{\smallskip}   
					&				&Freq.\supa	&$n_\mathrm{crit}$\supb &$E_\mathrm{u}$\supc &$v_\mathrm{disc}$\supd  &$T_\mathrm{disc}$\supe &$T_\mathrm{outf}$\supe 	&$N_\mathrm{disc}$\supe 	&$N_\mathrm{outf}$\supe	&			&			&			&	\\
Molecule				&Transition		&(MHz)		&(cm$^{-3}$) &(K)	&(km/s)		&(K)		&(K)			&(cm$^{-2}$)			&(cm$^{-2}$)				&$\tau_\mathrm{disc}$\supd &$\tau_\mathrm{outf}$\supd &$X_\mathrm{disc}$\supd	&$X_\mathrm{outf}$\supd	&$\frac{X_\mathrm{outf}}{X_\mathrm{disc}}$	\\
\hline
\noalign{\smallskip}
$^{13}$CO			&2 -- 1				&220398.7	&$6.0\times10^3$&16	&$-6.4$		&100		&100			&$7.9\times10^{16}$		&$2.6\times10^{17}$			&0.04	&0.23	&$2.9\times10^{-8}$		&$5.0\times10^{-7}$	&17			\\
SO~$^3\Sigma$		&6$_5$ -- 5$_4$		&219949.4	&$7.1\times10^5$&35	&$-3.4$		&230		&180			&$3.4\times10^{16}$		&$1.3\times10^{16}$			&0.42	&0.26	&$1.2\times10^{-8}$		&$2.5\times10^{-8}$	&2.0		\\
H$_2$$^{13}$CO		&3$_{1,2}$ -- 2$_{1,1}$	&219908.5	&$4.6\times10^5$&33	&$-2.4$		&230		&180			&$5.6\times10^{15}$		&$1.9\times10^{15}$ 		&0.07	&0.04	&$2.0\times10^{-9}$		&$3.6\times10^{-9}$	&1.7		\\
HNCO				&10$_{0,10}$ -- 9$_{0,9}$&219798.3	&$8.6\times10^5$&58--102	&$-3.4$		&230		&180			&$5.2\times10^{16}$		&$1.0\times10^{16}$ 		&0.37	&0.13	&$1.9\times10^{-8}$		&$2.0\times10^{-8}$	&1.0		\\
\hline
CH$_2$CO			&11$_{1,11}$ -- 10$_{1,10}$&220178.2	&$-$&76	&$-2.4$		&230		&180			&$1.5\times10^{16}$		&$5.5\times10^{15}$ 		&0.07	&0.04	&$5.6\times10^{-9}$		&$1.0\times10^{-8}$	&1.9		\\
$t$-HCOOH			&10$_{0,10}$ -- 9$_{0,9}$&220038.1 	&$-$&59	&$-1.4$		&230		&180			&$4.3\times10^{16}$		&$4.0\times10^{15}$ 		&0.12	&0.02	&$1.6\times10^{-8}$		&$7.5\times10^{-9}$	&0.5	\\
NH$_2$CN			&11$_{1,10}$ -- 10$_{1,9}$&221361.2	&$-$&63--295	&$-1.4$		&210(30)	&$-$			&$3.3\times10^{15}$		&$-$ 					&0.10	&$-$		&$1.2\times10^{-9}$		&$-$		&$-$				\\
\hline
\methanol\supf  	 	&10$_{-5,5}$ -- 11$_{-4,8}$&220401.4	&$3.3\times10^4$&252--802&$-0.4$		&230		&180			&$2.5\times10^{18}$		&$7.2\times10^{17}$ 		&0.21	&0.09	&$9.3\times10^{-7}$		&$1.4\times10^{-6}$	&1.5		\\
\methylcyanide\supg		&12$_{3}$ -- 11$_{3}$&220709.0		&$1.9\times10^6$&69--419&$-2.3$		&212(30)	&235(15)		&$2.1\times10^{16}$		&$3.8\times10^{15}$ 		&0.94	&0.14	&$7.7\times10^{-9}$		&$7.2\times10^{-9}$		&0.9	\\
\methylcyanide\,$_{v_{8}=1}$\suph &12$_{-6}$ -- 11$_{-6}$&221196.7&$-$&588--931&$-1.4$	&230(30)&186(15)	&$6.3\times10^{16}$		&$1.3\times10^{16}$ 		&0.21	&0.03	&$2.3\times10^{-8}$		&$2.4\times10^{-8}$		&1.0	\\
\hline
\methylformate\			&18$_{14,5}$ -- 17$_{14,4}$&221047.8	&$-$&103--357&$-1.4$		&250(30)	&170(15)		&$2.5\times10^{17}$		&$5.5\times10^{16}$			&0.04	&0.02	&$9.3\times10^{-8}$		&$1.0\times10^{-7}$	&1.1		\\
\ethylcyanide			&25$_{2,24}$ -- 24$_{2,23}$&220660.9	&$-$&143	&$-3.4$		&230		&180			&$5.2\times10^{15}$		&$1.2\times10^{15}$ 		&0.06	&0.02	&$1.9\times10^{-9}$		&$2.3\times10^{-9}$		&1.2	\\
\dimethylether			&24$_{4,20}$ -- 23$_{3,21}$&220893.1	&$-$&274--381&$-1.4$		&230		&180			&$1.0\times10^{17}$		&$8.1\times10^{16}$ 		&0.04	&0.04	&$3.9\times10^{-8}$		&$1.5\times10^{-7}$		&4.0	\\
\acetone\				&22$_{0,22}$ -- 21$_{0,21}$&220361.9	&$-$&63--310&$-1.4$		&230(30)	&190(20)		&$2.6\times10^{17}$		&$1.7\times10^{16}$ 		&0.11	&0.02	&$9.7\times10^{-8}$		&$3.2\times10^{-8}$		&0.3	\\
$_\mathrm{aGg'}$\ethyleneg\	&21$_{4,18}$ -- 20$_{4,17}$\supi&221007.8&$-$&122--156&$-1.4$		&230		&$-$			&$7.9\times10^{16}$		&$-$ 					&0.06	&$-$		&$2.9\times10^{-8}$		&$-$				&$-$		\\
\hline
\end{tabular}
\begin{list}{}{}
\item[$^\mathrm{a}$] Frequency of the transition imaged in Figs.~\ref{fm0simple} and \ref{fm0coms}, or frequency with highest opacity within the observed frequency range otherwise. In all cases the frequency is the nominal frequency for the corresponding transition as given in the JPL or CDMS databases (Picket et al. 1998, M\"uller et al. 2005), whose quantum numbers are given in column (2).
\item[$^\mathrm{b}$] Critical density at 200~K for the transition specified in columns (2) and (3), calculated using the data available in the Leiden Atomic and Molecular Database (LAMDA, Sch\"oier et al. 2005), and following equation (4) of Shirley (2015). 
For HNCO and CH$_3$CN, we give the critical density after logarithmic interpolation in temperature.
For  the case of \formald, we give the critical density for the 3$_{1,2}$ -- 2$_{1,1}$ transition of the main isotope reported by Shirley (2015), which corresponds to a temperature of $\sim100$~K.
\item[$^\mathrm{c}$] Upper level energy corresponding to the transition specified in columns (2) and (3). For those molecules for which we detected more than one transition (see Appendix A), we give the range of upper level energies covered by the different detected transitions (at the outflow position, or at the disc position if the molecule was only detected in the disc).
\item[$^\mathrm{d}$] Velocity adopted for each molecule (see Section 3) at the disc position. The velocity adopted at the outflow position is $-3.4$~\kms\ for all molecules except for $^{13}$CO and CH$_3$CN, for which we adopted -6.4 and -4.4~\kms, respectively. 
\item[$^\mathrm{e}$] $T$ and $N$ correspond to the rotational temperature and column density resulting from the XCLASS fit for the disc and outflow positions. These values are also the ones used to build the synthetic spectra shown in Fig.~\ref{fspec} and Appendix A. 
Wherever $T$ is given with a number in parenthesis (the uncertainty), it has been fitted by XCLASS. If no number in parenthesis is given, $T$ has been fixed to the average value for the disc and outflow positions. 
$\tau$ corresponds to the opacity of the transition given in {\bf columns (2) and (3)}. The abundances, $X$, are given relative to the column density of H$_2$ (see Section~\ref{res}). 
\item[$^\mathrm{f}$] Blended with $^{13}$CO (see Figs. A1 and A8 of Appendix A).
\item[$^\mathrm{g}$] \methylcyanide\ has been fitted using two components: a cold foreground component, and a hot inner component. In this table we list only the parameters corresponding to the hot component.
\item[$^\mathrm{h}$] Transition with a small contribution from \dimethylether\ (see Figs. A10 and A13 of Appendix A).
\item[$^\mathrm{i}$] This transition corresponds to $v=1$ -- $v=0$.
\end{list}
\label{tcoms}
\end{table}
\end{landscape}

\begin{table}
\caption{Morphological properties of the emission of complex organic molecules from \src\ at a spatial resolution of $\sim600$~au  }
\centering
\begin{tabular}{l c c c c}
\hline\hline\noalign{\smallskip}   
					&Ang. size\supb	&P.A.\supb	&Size\supb			\\
Molecule\supa			&($''\times''$)	      	&(\degr)        	&(au$\times$au)		\\
\hline
\noalign{\smallskip}
H$_2$$^{13}$CO		&$0.74\times0.49$  	&\phe$75\pm10$	&$1210\times800$\phe  		\\
HNCO				&$0.53\times0.46$ 	&$114\pm11$		&$870\times750$		\\
CH$_2$CO			&$0.54\times0.41$ 	&$102\pm20$		&$890\times670$		\\
HCOOH				&$0.48\times0.37$ 	&$105\pm35$		&$790\times610$		\\
\methanol\supc  	 	&$0.83\times0.79$  	&\phe$85\pm8$\phe	&$1360\times1300$ 		\\
\methylcyanide\,$v_8$=1	&$0.55\times0.42$  	&\phe$88\pm5$\phe	&$720\times610$		\\
\methylformate			&$0.73\times0.43$ 	&\phe$98\pm37$	&$1200\times710$\phe  		\\
\ethylcyanide			&$0.56\times0.50$ 	&\phe$91\pm42$	&$920\times820$		\\
\dimethylether			&$0.69\times0.49$  	&\phe$96\pm9$\phe	&$1130\times800$\phe  		\\
\acetone\				&$0.52\times0.43$  	&\phe$96\pm44$	&$850\times710$		\\
\hline
\end{tabular}
\begin{list}{}{}
\item[$^\mathrm{a}$] The frequency and quantum numbers of the imaged transition is given in Table~\ref{tcoms}.
\item[$^\mathrm{b}$] Deconvolved sizes and P.A. obtained from a 2D-Gaussian fit to the zeroth-order moment image of each transition. Uncertainties are typically $\la0.1$~arcsec, except for \methylformate\ and \ethylcyanide, which are $\sim0.2$--0.3~arcsec.
\item[$^\mathrm{c}$] Blended with $^{13}$CO (see Figs. A1 and A8 of Appendix A).
\end{list}
\label{tsizes}
\end{table}

\begin{figure*}
\begin{center}
\begin{tabular}[b]{c}
    \epsfig{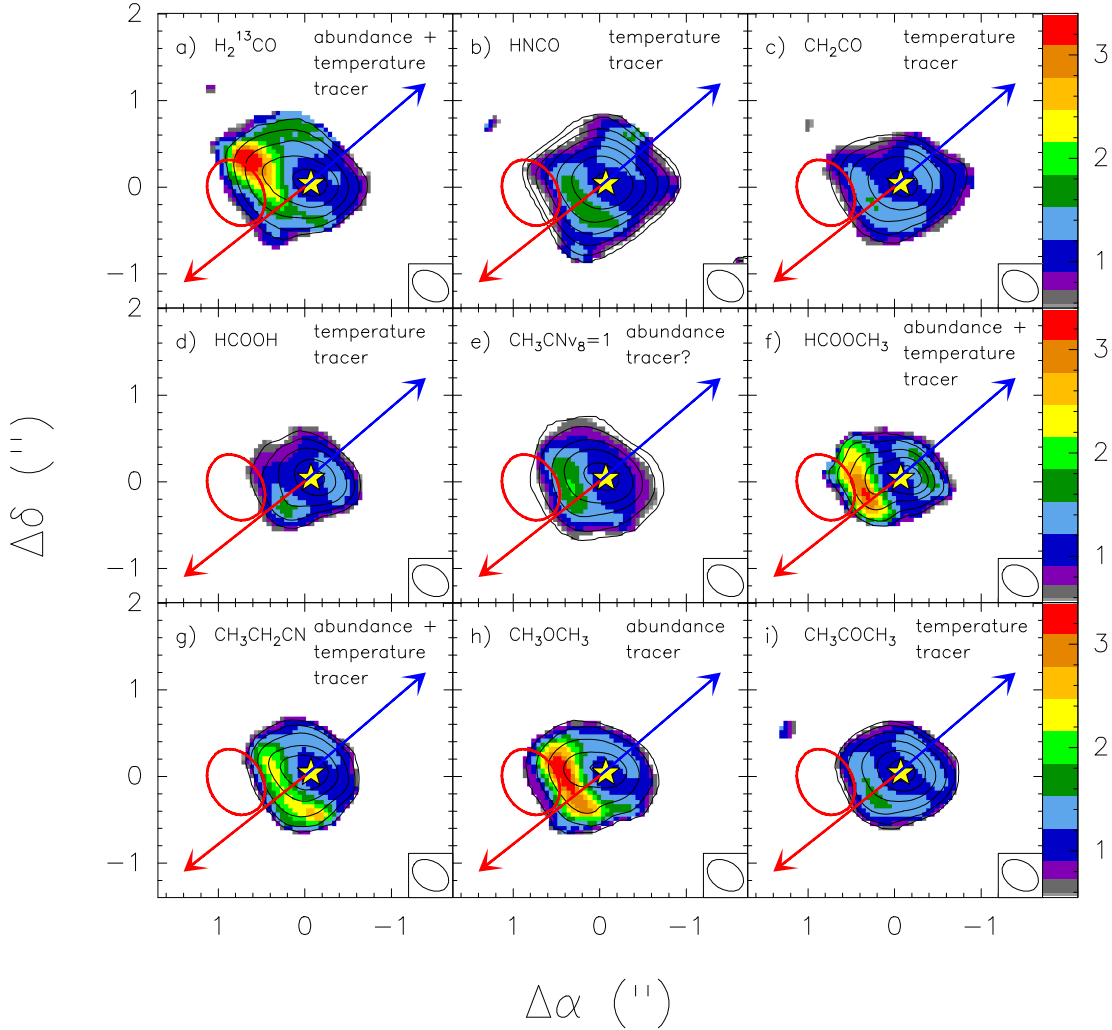}\\
\end{tabular}
\caption{Normalised line-to-continuum ratio maps (generated by computing the ratio of the PdBI zeroth-order moment maps from Fig.~\ref{fm0coms} with respect to the 1~mm continuum emission, relative to the value of the ratio at the 1~mm peak so that all figures show a comparable scale). The star symbol marks the position of the 1.4~mm peak from Cesaroni et al. (2014, also shown in Fig.~\ref{fm0simple}a), and the red ellipse indicates the H$_2$ knot tracing the outflow cavity (Cesaroni \et\ 2013, marked also in Fig.~\ref{fm0coms}a). See Table~\ref{tF} for further details about the physical conditions (abundance/temperature) which each line-to-continuum ratio map is probably tracing.
\vspace{2cm}
}
\label{fratios}
\end{center}
\end{figure*}

\section{Analysis}\label{ana}

In this Section, we aim at studying more deeply the origin of the COM emission and the abundance enhancement by comparing the continuum vs the line emission maps, and by running a chemical model to test different hypotheses for the chemical origin of COMs.

\subsection{Line-to-continuum ratio maps}\label{mod}

In the previous section we found that the abundance of COMs at the outflow position is of the same order as the abundance of COMs at the position of the disc. Such a high abundance at the outflow position may be due to either a real enhancement of COMs at this position, or to uncertainties in the excitation temperature, as an underestimation of the excitation temperature at the outflow position would yield a higher column density.
By performing line-to-continuum ratio maps, in this section we evaluate which COMs have been enhanced truly by the passage of the outflow.

The 1.4~mm continuum emission shown in Fig.~\ref{fm0simple}-a presents an elongation towards the south-east, which is coincident with a H$_2$ infrared knot tracing the outflow cavity (Sridharan \et\ 2005; Cesaroni et al. 2013). 
Similarly, many of the partially resolved COMs in \src\ also present an elongation towards the south-east, suggesting that such an elongation might be associated with the outflow cavity as well. 
Since, at first order, the integrated intensity maps trace the COMs column density, and the continuum map traces the total gas column density, with their ratio being a first approach to the abundance, we computed the ratio of the integrated intensity maps of different COMs to the continuum emission (normalized to the value of that ratio at the peak of the continuum emission, which is the putative position of the disc), shown in Fig.~\ref{fratios}.

Fig.~\ref{fratios} reveals for several COMs an enhancement in the normalized line-to-continuum ratio towards the (south)east of \src, \ie\ along the outflow direction. This enhancement is particularly strong in the case of \formald, \methylformate, and most importantly \dimethylether\ (see Fig.~\ref{fratios}a, f, h), with the ratio of the COM in the cavity wall being more than three times higher than the ratio in the disc. In particular, for the two molecules for which this ratio is highest (\formald\ and \dimethylether), the normalized line-to-continuum ratio peaks about $\sim0.5$~arcsec to the east of the peak of the continuum, and lies exactly upstream (to the west) of the H$_2$ knot reported by Cesaroni \et\ (2013), interpreted as tracing the outflow cavity wall (marked with a red ellipse in Figs.~\ref{fm0coms}a and \ref{fratios}). 

We consider now which physical process the line-to-continuum ratio may be tracing. In the case of optically thin emission, and assuming the Rayleigh-Jeans approximation, that the background temperature is much lower than the excitation temperature, and that the dust opacity is similar at the disc and outflow positions, the line-to-continuum ratio, normalized to the ratio at the disc, can be expressed as:

\begin{equation}
\frac{\bigg[\frac{T_\mathrm{B,line}}{T_\mathrm{B,cont}}\bigg]_\mathrm{outf}}{\bigg[\frac{T_\mathrm{B,line}}{T_\mathrm{B,cont}}\bigg]_\mathrm{disc}} = 
\frac{X_\mathrm{outf}}{X_\mathrm{disc}}
\frac{Q_{T_\mathrm{ex,disc}}\,T_\mathrm{dust,disc}\,e^{E_\mathrm{u}/(k\,T_\mathrm{ex,disc})}}{Q_{T_\mathrm{ex,outf}}\,\,T_\mathrm{dust,outf}e^{E_\mathrm{u}/(k\,T_\mathrm{ex,outf})}},
\end{equation}

where $T_\mathrm{B}$ is the brightness temperature of the line or the continuum emission, $X$ is the abundance of the molecule, $Q$ is the partition function at a given excitation temperature $\Tex$, $E_\mathrm{u}$ is the energy of the upper level of the imaged transition, $k$ is the Boltzmann constant, and $\Tdust$ is the dust temperature. 
In our case, all the molecules for which the line-to-continuum ratio is calculated present optically thin emission ($\tau\la0.4$) even at the putative position of the disc (see Table~\ref{tcoms}).
Therefore, the normalized line-to-continuum ratio depends essentially on the different abundances of the molecule at the disc and outflow positions, and on the different $\Tex$ and $\Tdust$ at the disc and outflow positions. 

We define a factor $F$ which includes the dependence of the normalised line-to-continuum ratio on $\Tex$, $\Tdust$, and $E_\mathrm{u}$:

\begin{equation}
F\equiv\frac{Q_{T_\mathrm{ex,disc}}\,T_\mathrm{dust,disc}\,e^{E_\mathrm{u}/(k\,T_\mathrm{ex,disc})}}{Q_{T_\mathrm{ex,outf}}\,T_\mathrm{dust,outf}\,e^{E_\mathrm{u}/(k\,T_\mathrm{ex,outf})}}.
\end{equation}

Taking into account that the dependence of the partition function with $\Tex$ for symmetric and asymmetric top molecules is proportional to $T^{3/2}$, and assuming that $\Tex\,\sim\,\Tdust$, $F$ can be finally written as:

\begin{equation}
F = \bigg(\frac{T_\mathrm{disc}}{T_\mathrm{outf}}\bigg)^{5/2}\,e^{\frac{E_\mathrm{u}}{k}\,(1/T_\mathrm{disc}-1/T_\mathrm{outf})}.
\end{equation}

We note that the assumption $\Tex\,\sim\,\Tdust$ should be valid as a first approach because what is relevant to estimate $F$ is the ratio of those temperatures at the disc and outflows positions, and it is likely that both temperatures decrease in a similar way from one position to the other. We further discuss this assumption in Section~5.1.

Thus, if $F\sim1$, the normalized line-to-continuum ratio will trace the abundance variation between the disc and outflow positions. But if $F$ is of the order of the normalized line-to-continuum ratio, this ratio will be a temperature tracer rather than an abundance tracer. Since we have modeled the emission of the detected molecules at the disc and outflow positions using XCLASS, we can have a first estimate of $F$ for the transitions for which we have imaged the line-to-continuum ratio. The values of $F$ for these transitions are reported in Table~\ref{tF}, along with the observed peak of the normalized line-to-continuum ratio.

\begin{table}
\caption{Parameters used to assess the relevance of the abundance vs temperature in the line-to-continuum ratio for certain transitions of COMs detected in \src.  }
\centering
\begin{tabular}{l c c c c}
\hline\hline\noalign{\smallskip}   
					&$E_\mathrm{u}$\supb				\\
Molecule\supa			&(K)	      				&$F$\supc	&Ratio\supd	&Tracing?\supe\\
\hline
\noalign{\smallskip}
\formald\				&\phe32.9  			&1.77		&3.4			&A+T\\
HNCO				&\phe58.0 			&1.72		&1.6			&T\\
CH$_2$CO			&\phe76.5 			&1.68		&1.6			&T\\
HCOOH				&\phe58.6 			&1.72		&1.6			&T\\
\methylcyanide\,$v_8$=1	&931.0  				&0.65		&2.0			&A?\\
\methylformate			&230.9 				&1.78		&3.2			&A+T\\
\ethylcyanide			&143.0 				&1.55		&2.4			&A+T\\
\dimethylether			&274.4  				&1.32		&3.2			&A\\
\acetone\				&123.9  				&1.44		&1.6			&T\\
\hline
\end{tabular}
\begin{list}{}{} 
\item[$^\mathrm{a}$] The frequency and quantum numbers of the imaged transition is given in Table~\ref{tcoms}.
\item[$^\mathrm{b}$] Energy of the upper level of the imaged transition.
\item[$^\mathrm{c}$] $F$ as defined in equation (3), using the temperatures for the disc and outflow positions reported in Table~\ref{tcoms}.
\item[$^\mathrm{d}$] Maximum normalized line-to-continuum ratio as measured in the maps shown in Fig.~\ref{fratios}.
\item[$^\mathrm{e}$] Tentative indication of what the line-to-continuum ratio might be tracing for each transition, after comparing the observed line-to-continuum ratio given in column (4) with the contribution of the different disc and outflow temperatures assessed by the `F' factor given in column (3). `A' for `Abundance'; `T' for `Temperature'.
\end{list}
\label{tF}
\end{table}

Table~\ref{tF} shows that the observed normalized line-to-continuum ratio is very similar to the $F$ factor for HNCO, CH$_2$CO, HCOOH, and \acetone, indicating that the variations in the line-to-continuum map might be due to different temperatures in these cases. On the contrary, for the case of \formald, \dimethylether, and, to a less extent, \methylformate\footnote{Strictly speaking, also \methylcyanide\,$_{v_8=1}$ presents a large difference between the observed normalized line-to-continuum ratio and the $F$ factor, but we avoid its discussion here because the modelling of \methylcyanide\ required two components, making its interpretation more difficult.}, the observed ratio is still larger by more than a factor of $\sim2$ compared to the $F$ factor, and thus the observed variations in the normalized line-to-continuum maps could be indicative of a real abundance enhancement at the position of the H$_2$ knot tracing the outflow cavity. This is also consistent with the ratio of the abundance at the outflow versus the abundance at the disc obtained from the XCLASS fit for \formald, and \dimethylether\ (Table~\ref{tcoms}).


Thus, our data suggest that for those COMs for which the line-to-continuum ratio is sensitive to abundance variations, the abundance of these molecules peaks at the position of the H$_2$ knot. This, together with the result presented in the previous section that the abundances of most COMs at the disc and outflow positions are comparable, 
and the fact that the gas temperature at the outflow position is lower than in the disc, suggests a mechanism, additional to thermal desorption, of enhancing the COM abundances in the gas phase, probably related to the passage of a shock. 
In the next section we aim at studying the different processes that might be at work in the outflow cavity to efficiently enhance the abundance of COMs, namely shocks and UV radiation from the central massive protostar.

\subsection{Chemical modeling}\label{mod}

We first studied the possibility that COMs are processed and released into the gas phase via photo-processes triggered by the stellar radiation field. To do this, we ran models over a grid of $A_\mathrm{v}$ (from 1 to 100 mag) to simulate different depths along the outflow cavity. 
In these models, presented in Walsh et al. (2010, 2012, 2014), density, impinging flux, gas and dust temperature, and visual extinction are kept constant.
We find that for $A_\mathrm{v} < 10$ mag most species cannot survive the strong radiation field in the gas phase (G$\sim10^5$ Gism, in Draine units, Draine 1978), with the exception of H$_2$ and CO which can self-shield, and the grains are also stripped of the ice mantle. 
Further into the cavity wall, the solution tends towards the `high extinction' results, where thermal desorption dominates.  In regions where the photodesorption rate is faster than the thermal desorption rate, the photodissociation rates are also high so that molecules released from the grains are destroyed in the gas phase. 
Furthermore, recent photodesorption experiments also show that larger molecules do not photodesorb intact (Bertin et al. 2016; Cruz-Diaz et al. 2016).
Therefore, the UV radiation from the massive protostar does not seem to favor the enhancement of COM abundances in the gas phase, and in the following we consider the role of shocks along the outflow cavity.

To test our hypothesis that shocks may play a role in the liberation of COMs from the grain mantle in the outflow cavity walls, we calculate the chemical evolution of the gas and grain mantle as a function of time (or distance) from the shock front.  We assume a C-type shock with shock velocity of 40~\kms, a pre-shock density of $10^4$~\cmt, and a pre-shock gas and dust temperature of 80 K.  
The shock physical structure was constructed following the parametric approximation of Jim\'enez-Serra et al. (2008) for C-type shocks. 
The value of the pre-shock dust temperature is taken from the modeled temperature profile of \src\ at 1000~AU (Palau et al. 2014), and the value of the density corresponds to an average of the typical values estimated in outflow cavity walls (\eg\ Neufeld et al. 2009, Bruderer et al. 2009, Visser et al. 2012, Gomez-Ruiz et al. 2015, Kuiper et al. 2015, Gusdorf et al. 2016).

\begin{figure}
\begin{center}
\begin{tabular}[b]{c}
    \epsfig{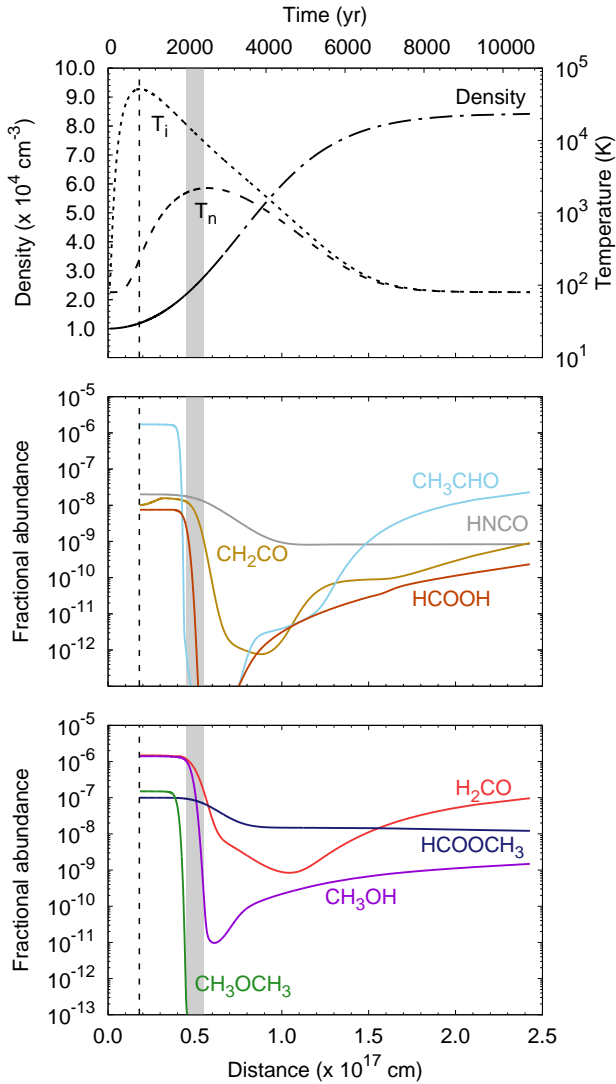}\\
\end{tabular}
\caption{
\emph{Top:} Physical structure of a shock with velocity, $V_\mathrm{s} = 40$~\kms, and a pre-shock density of $10^4$~\cmt.  $T_\mathrm{i}$ and $T_\mathrm{n}$ represent the temperature of the ions and neutrals, respectively.
\emph{Middle and bottom:} Fractional abundances (with respect to H$_2$) of gas-phase COMs as a function of distance from the shock front. In all panels, the vertical dashed line corresponds to the peak ion temperature (see text for details). The thick grey vertical line indicates the timescale when most of the molecules keep their initial abundance, of $\sim2000$~yr.
\vspace{2cm}
}
\label{fmodel}
\end{center}
\end{figure}


Concerning the shock velocity, a water maser spot is detected to the south-east of \src\ with about 15~\kms\ (plane-of-sky velocities, Cesaroni et al. 2014), while radial velocities are detected up to $\sim25$~\kms\ (Cesaroni et al. 1999, Cesaroni et al. 2005, this work), which makes a total of 
$\sim30$~\kms. Adopting a shock velocity slightly higher (40~\kms) seems reasonable because there is only one water maser to the south-east of \src, and the large number of water masers found to the north-west suggest a higher velocity.


We assume that the gas and dust temperatures are decoupled, with the dust remaining cold (80 K).  We begin the chemical evolution calculation at the time step corresponding to the peak temperature of the ions which also corresponds to the peak sputtering rate. We assume at this point that all grain mantle material is sputtered into the gas phase.  For the COMs observed in \src\ we assume the initial abundance is equal to that observed (Table~\ref{tcoms}), while for all other species we adopt initial abundances from a single-point molecular cloud model (10 K, $10^4$~\cmt). 
For H$_2$CO, we also use the model output, as opposed to converting the abundance from H$_2$$^{13}$CO to H$_2$$^{12}$CO, because H$_2$CO has gas-phase pathways to formation, unlike many of the larger COMs (e.g., CH3OH) which likely originate mainly (or solely) from the ice mantle.
The abundances measured in \src\ are about one-two orders of magnitude larger than the abundances measured in pre-stellar cores (Vastel et al. 2014, Jim\'enez-Serra et al. 2016), which is reasonable given the hot core nature and the size scales (much closer to the massive protostar) that we are studying here, and are a factor of 3--10 larger than the abundances measured by Taquet et al. (2015) toward the hot corino NGC\,1333-IRAS2A at similar spatial scales.

The chemical network used here is based on the latest release of the UMIST Database for Astrochemistry (RATE12, McElroy et al. 2013) supplemented with gas-grain interactions (freeze out and desorption) and grain surface chemistry.  The grain-surface network and associated gas-phase chemistry is extracted from the Ohio State University (OSU) network (Garrod et al. 2008).  The reaction rate coefficients are calculated as described in Walsh et al. (2012, 2014).  The gas-phase chemistry was also supplemented with additional neutral-neutral reactions extracted from the NIST Chemical Kinetics Database (http://kinetics.nist.gov/kinetics/index.jsp).  These reactions included collisional dissociation and association and reactions with atomic H and O and small radicals such as CH$_3$ and OH (see Appendix B).

The shock structure and chemical evolution results are shown in Figure~\ref{fmodel}.  The COMs survive in the gas-phase until a time of $\sim2000$~years when they are destroyed by reactions with atomic hydrogen, small radicals such as OH, and collisional dissociation (corresponding to a temperature of $\sim2000$~K for the neutral gas). 
Thus, the main effect of the shock is the increase in temperature, which favors the aforementioned reactions as they have a significant barrier.
The timescale for destruction for two-body reactions, if the temperature is high enough, is set by the density.  The peak in the sputtering rate in our model corresponds to only 200--300 K in the neutral species (gas is mainly neutral), so the gas-phase species do not get destroyed until the neutral temperature approaches its peak of about 2000~K.  Thus, the timescale of $\sim2000$~yr is a combination of both density and the rise time of the neutral gas temperature post shock.
The key point of our model is that it shows that COMs can survive long enough after the passage of a shock.
As the temperature falls in the post shock regime, several COMs begin to increase in abundance again, mainly through reactions in the gas phase,  except for \methanol\ for which some grain-surface chemistry contributes.  
\dimethylether\ does not recover at all, because it is solely reliant on grain-surface reformation.
There are two molecules, \methylformate\ and HNCO, which remain fairly flat with time. This is because their collisional dissociation rates are significantly slower at the peak of the gas temperature (by at least an order of magnitude) than those for the other molecules.

The simple model presented in Section 4.2 shows that COMs such as the ones detected in \src\ (\eg\ \formald, \methylformate, \dimethylether) can survive in the gas phase for a sufficient period following the passage of a C-type shock of about 40~\kms.  For slower shocks, the peak temperature of the ions and neutrals gets lower, impeding the destruction of COMs via neutral-neutral reactions; however, sputtering of the grain mantle is less efficient.  For higher density material, the timescale of the shock and the chemistry is shorter, e.g., for a density of $10^5$~\cmt, the COMs survive in the gas for $\sim200$ years.

\section{Discussion}\label{dis}

In previous sections we presented subarcsecond angular resolution images of several COMs associated with \src. 
In almost all cases the morphology of the COMs presents an elongation towards the south-east, \ie\ along the outflow direction.
For most of the COMs (with the exception of \acetone) our XCLASS fits to the spectra at the disc and outflow positions reveal that the abundance at the outflow position is comparable or even larger than the abundance at the disc. We performed normalised line-to-continuum ratio maps in order to locate the abundance peak in the surroundings of \src\ and found that for \formald\ and \dimethylether\ the abundance peaks at the position of an H$_2$ knot rather than at the disc position (Fig.~\ref{fratios}).

The fact that \dimethylether\ is among the COMs showing the strongest abundance enhancement along the outflow direction is consistent with observations towards other high-mass star-forming regions. In particular, \dimethylether\ 
is found in OrionKL further to the south compared to other COMs (Favre et al. 2011a, 2011b, Peng et al. 2013), suggesting its association with shocks. Similarly, 
\"Oberg et al. (2011) report the detection of  \dimethylether\ towards the shock SMM4-W near a low-mass protostar.

This suggests that the mechanism which enhances the emission from COMs may not be exclusively related to the presence of a massive protostar, but could also be triggered by the passage of a shock.
The chemical modelling coupled to a parametric shock model presented in Section 4.2 shows that COMs like CH$_2$CO, HNCO, \methylformate, \dimethylether, and H$_2$$^{13}$CO can survive in the gas phase for $\sim2000$~yr after the passage of a C-type $\sim40$~\kms\ shock. 

If a non-negligible amount of COMs in \src\ result from shocks along the outflow, COMs should present relatively broad linewidths. However, we could fit the spectra at the disc and outflow positions using line widths of about $\sim8$~\kms, for almost all the molecules (including the outflow tracer SO). 
This is suggesting that, at least at these two positions, most of the COMs are not coming from the high velocity gas itself but from the warm cavity walls, or from a post-shock region where the gas has been already decelerated.

The current models for the formation of COMs consider a dense and cold phase where simple hydrogenated species are formed on the surface of dust grains, followed by a warm-up phase where efficient recombination of radicals takes place on the grain surfaces (\eg\ Garrod \& Herbst 2006; Garrod et al. 2008; Aikawa et al. 2008). However, these models underpredict the abundance of \methylformate\ and \dimethylether\ with respect to their parent molecule \methanol\  (\eg\ Figure 1 of Taquet et al. 2012).
Our observations suggest that additional ingredients related to shocks should be taken into account by these models.
For example, COMs could be produced through ion-neutral gas phase chemistry in the post-shock region for moderate temperatures (of $\sim100$~K, as these are temperatures for which COMs begin to increase in abundance again following the passage of the shock). 
Taquet et al. (2016) re-investigated the formation of several COMs, such as \methylformate\ and \dimethylether, for hot core conditions 
and found that ion-neutral chemistry (in which proton-transfer reactions involving ammonia play a key role) can produce COMs up to abundances of $10^{-6}$. It remains to be investigated this gas phase chemistry for shock conditions with lower densities and shorter timescales.
Another ingredient which might help models recover the observed abundances is sputtering of the grain mantles by the passage of a shock. This is actually supported by the fact that  the weakly bound molecules (\eg\ H$_2$CO, HNCO, CH$_2$CO) are co-spatial with the strongly bound COMs (larger COMs such as HCOOCH$_3$, CH$_3$OCH$_3$...), pointing to a global mechanism, additional to thermal desorption, which is releasing them to the gas phase.


\subsection{Caveats}

A possible caveat in our derivation of abundances is that if the density were not high enough to thermalize dust and gas in the outflow, then the dust temperature could be lower than the gas temperature. In this case the total H$_2$ column density would have been underestimated and the abundances of the molecules would have been overestimated. Since the density in the outflow cavity should be lower than in the disc, this should mostly affect the abundances derived at the outflow position. Palau et al. (2014) estimate the dust temperature to be $\sim80$~K at the radius corresponding to the outflow position\footnote{The estimate of 80~K as the dust temperature at the outflow position is a first-order approach and probably a lower limit, as this temperature is inferred from a spherical model which does not consider the presence of cavities produced by outflows (Palau et al. 2014).}, of $\sim1000$~au. If we recalculate the abundances at the outflow position assuming such a dust temperature in the estimate of the H$_2$ column density, the abundances decrease only by a factor of two. Thus our conclusion that the abundances of COMs at the outflow position are of the same order, or might be even higher in some cases, as the abundances at the disc position is robust. 

It is also important to keep in mind that in the model presented in Section~\ref{mod} the assumed pre-shock densities ($\sim10^4$~\cmt) might be low given the embedded and massive nature of \src. As explained above, if we used a pre-shock density of $\sim10^5$~\cmt, COMs would survive in the gas phase about $\sim200$~yr. Such short timescales for the COM enhancement are consistent with the timescales of shocked near-infrared H$_2$ emission, clearly spatially associated with the COM enhancement, of $<1000$~yr (Gusdorf et al. 2011), which supports the idea that we are seeing a `short' time-dependent phenomenon. 
Furthermore, this timescale is of the order of the approximated lifetime of water masers tracing the outflow at this position (Cesaroni et al. 2013, 2014), as seen also in other sources (\eg\ Burns et al. 2016, Burkhardt et al. 2016), and is comparable to the timescale required by models used to explain the SiO or \acetaldehyde\ emission in shocks (Gusdorf et al. 2011, Codella et al. 2015). 
%
In addition, these short timescales could also explain why no COM enhancement is detected in the northern outflow cavity.
Therefore, the short timescales required by our model seem feasible, supporting the idea that the COMs detected along the outflow in \src\ have been evaporated from the dust grains by the shocks associated with the outflow.

We end with the caveat that, despite the large extension to the high-temperature network described in Appendix B to run the model presented in Section 4.2, such a network likely remains incomplete for many gas-phase COMs given the lack of data in the literature.  
If our theory is correct, \ie\ that gas-phase COMs in intermediate- to high-mass protostars originate from shock-induced sputtering instead of the traditional `hot core', then we would urge 
the gas-phase laboratory community to revisit high-temperature gas-phase chemistry with a view to quantifying the chemistry of COMs in post-shocked gas.  
Future high angular ($\sim$~0.1~arcsec)  resolution observations (with e.g., ALMA and NOEMA) towards a broad sample of sources will help further elucidate the origin of COMs in protostars.

\section{Conclusions}

We present subarcsecond angular resolution observations carried out with the PdBI at 220.5~GHz towards \src. These observations reveal for the first time resolved emission from COMs towards \src\ with a spatial resolution of $\sim600$~au. Our main conclusions are summarized as follows:

\begin{itemize}

\item We detect and resolve the emission from SO, HNCO, \formald, \methylcyanide, \methanol, HCOOH, \methylformate, \dimethylether, \ethylcyanide, and \acetone. In addition, we detect NH$_2$CN and \ethyleneg. 

\item While SO and HNCO present a X-shaped morphology consistent with tracing the outflow cavity walls, most of the COMs have their peak emission at the putative position of the protostar (the disc), but also show an extension towards the south(east), coinciding with an H$_2$ knot from the outflow at about 800--1000~au from the protostar. This is especially clear for the case of \formald, HCOOH, \methylformate, and \dimethylether.

\item By fitting the observed spectra at two representative positions for the disc and the outflow, we found that the abundances of most COMs are comparable at the two positions. In the case of \formald\ and \dimethylether\ the abundace at the outflow position is larger by a factor of 2--4 than the abundance in the disc. This suggests a mechanism of enhancing the COM abundances related to the passage of a shock.

\item We explored the possibility that the COMs detected along the outflow direction are enhanced by the UV-radiation from \src\ escaping through the outflow cavity wall. However, for $A_\mathrm{v} < 10$ mag most species cannot survive the strong radiation field in the gas phase. 

\item We coupled a parametric shock model to a large gas-grain chemical network including COMs, and find that the observed COMs should survive in the gas phase for $\sim2000$~yr, comparable to the shock lifetime estimated from the water masers at the outflow position.

\end{itemize}

Therefore, our data indicate that COMs in \src\  may arise not only from the disc, but also from dense and hot regions associated with the outflow.


\section*{Acknowledgments}
The authors are grateful to the anonymous referee, for his/her insightful comments which helped improve the clarity and quality of the paper.
A.P. is grateful to Vianney Taquet, Simon Bruderer, and Ewine van Dishoeck for thoughtful discussions.  
A.P. acknowledges financial support from UNAM-DGAPA-PAPIIT IA102815 grant, M\'exico.
C.W acknowledges support from the Netherlands Organisation for Scientific Research (NWO, program number 639.041.335).
A.S.M. is partially supported by the Collaborative Research Centre SFB 956, sub-project A6, funded by the Deutsche Forschungsgemeinschaft (DFG).
I.J.-S. acknowledges the financial support received from the STFC through an Ernest Rutherford Fellowship (proposal number ST/L004801/1).
A.F. thanks the Spanish MINECO for funding support from grants FIS2012-32096, AYA2012-32032, and ERC under ERC-2013-SyG, G. A. 610256 NANOCOSMOS.
L.A.Z. is grateful to CONACyT, M\'exico, and DGAPA, UNAM for their financial support.  
The authors acknowledge the efforts of the GILDAS team to keep this software up to date.

{}

\begin{appendix}

\section{Synthetic spectra}

In this Appendix we present the best-fit synthetic spectra for individual molecules generated using XCLASS with the parameters given in Table~\ref{tcoms}. [Contact the authors to receive a copy of the figures of this Appendix]

\clearpage

\section{Supplementary neutral-neutral reactions for COMs}

As stated in the main text, the gas-phase chemical network used in this work is the full  {\sc Rate}12 network (McElroy et al. 2013)\footnotemark\footnotetext{\url{http://www.udfa.net}} supplemented with gas-grain chemistry for COMs extracted from the Ohio State University (OSU) network (Garrod et al. 2008)\footnotemark\footnotetext{\url{http://faculty.virginia.edu/ericherb/research.html}}.  
Both networks originate from those created to describe relatively cold ($\sim10$~K) 
to warm ($\sim300$~K) ion-molecule chemistry; however, the philosophy of the UMIST Database for Astrochemistry 
in later releases was to also include chemistry applicable at higher temperatures ($\sim$~1000~K).  
In recent times, the OSU network was also expanded to include high-temperature chemistry 
which was then used to model the chemistry in the inner molecular disk of an active galactic nuclei 
(Harada et al. 2010, 2012). 
However, both networks concentrated on the neutral-neutral chemistry of small, simple molecules.  
Hence, current publicly available versions of both networks are lacking in high-temperature 
chemistry applicable for complex molecules, which is 
necessary to simulate the gas-phase chemistry of such species in 
post-shocked gas which can reach temperatures $\gg1000$~K. 

In Table~\ref{tableA1}, we list the reactions we have added in order to better 
describe the high-temperature neutral-neutral chemistry of gas-phase COMs.  
Most reactions have been extracted from the 
NIST Chemical Kinetics Database\footnotemark\footnotetext{\url{http://kinetics.nist.gov/kinetics/index.jsp}}  
and have their origin in either atmospheric chemistry or combustion chemistry studies 
(see also, Walsh et al. 2015).
We have restricted most of the reactions to those involving atomic H and O, and small radicals 
such as OH, \ce{CH3}, and HCO.  
There is a lack of information in the literature on reactions involving 
both atomic C and N; hence, we concentrate our efforts on the O-bearing COMs and 
associated radicals.  
Our approach was to adopt evaluated values of the rate coefficient where available.  
In several cases, a measured rate coefficient was adopted.  
We include theoretically derived rate coefficients only for those few reactions for which 
neither an evaluated nor a measured value was available.    
These latter values were also cross-referenced with other recent 
compilations in the literature (\eg\ Ruaud et al. 2015).  
We also include at least one collisional dissociation pathway for 
each COM, again, adopting an evaluated or measured rate coefficient 
where available.  
Often the collisional partners in such measurements are 
those applicable for atmospheric or combustion chemistry, e.g., 
\ce{N2} or \ce{CH4}.  
In some cases, a noble gas, e.g., Ar, is used.  
Given that the most likely collision partner in interstellar and 
circumstellar media is \ce{H2}, we crudely rescale all collisional rates 
by the ratio of the square root of the reduced mass of the measured system  to 
the \ce{H2} system ($k_\mathrm{col}\propto\sqrt{1/\mu_\mathrm{AB}}$ 
where $\mu_\mathrm{AB} = m_\mathrm{A}m_\mathrm{B}/(m_\mathrm{A}+m_\mathrm{B})$).  
The scaling factor is $\approx3$ for most reactions.  
For those species for which there is no data, e.g., \ce{HCOOCH3}, 
we adopt the approximation outlined in Willacy et al. (1998).


\newpage

\begin{footnotesize}
\begin{table*}
\caption{Reactions added to {\sc Rate}12.}
\begin{center}
{\small
\begin{tabular}{lcccc}
\noalign{\smallskip}
\hline\noalign{\smallskip}
              & $\alpha$ &         & $\gamma$      &  \\       
\multicolumn{1}{c}{Reaction}      &  (cm$^{3}\,$s$^{-1}$)              & $\beta$ & (K) & Reference \\ 
\noalign{\smallskip}
\hline\noalign{\smallskip}
$\ce{HNCO}    +\ce{H}  \longrightarrow\ce{NH2}    +\ce{CO}           $ & $5.00\times10^{-11}$ & 0.00 &   2300.0 & Tsang (1992) \\ 
$\ce{HNCO}    +\ce{H}  \longrightarrow\ce{OCN}    +\ce{H2}           $ & $1.46\times10^{-12}$ & 1.81 &   8330.0 & Tsang (1992) \\ 
$\ce{HNCO}    +\ce{O}  \longrightarrow\ce{OH}     +\ce{OCN}          $ & $6.24\times10^{-13}$ & 2.11 &   5750.0 & Tsang (1992) \\ 
$\ce{HNCO}    +\ce{O}  \longrightarrow\ce{CO2}    +\ce{NH}           $ & $4.98\times10^{-13}$ & 1.41 &   4290.0 & Tsang (1992) \\ 
$\ce{HNCO}    +\ce{OH} \longrightarrow\ce{H2O}    +\ce{OCN}          $ & $2.82\times10^{-13}$ & 1.50 &   1809.0 & Baulch et al. (2005) \\ 
$\ce{HNCO}    +\ce{OH} \longrightarrow\ce{CO2}    +\ce{NH2}          $ & $3.13\times10^{-14}$ & 1.50 &   1809.0 & Baulch et al. (2005) \\ 
$\ce{HNCO}    +\ce{HCO}\longrightarrow\ce{H2CO}   +\ce{OCN}          $ & $5.00\times10^{-12}$ & 0.00 &  13100.0 & Tsang (1992) \\ 
$\ce{CH3OH}   +\ce{H}  \longrightarrow\ce{CH2OH}  +\ce{H2}           $ & $3.36\times10^{-12}$ & 1.24 &   2260.0 & Baulch et al. (2005) \\ 
$\ce{CH3OH}   +\ce{H}  \longrightarrow\ce{CH3O}   +\ce{H2}           $ & $3.36\times10^{-12}$ & 1.24 &   2260.0 & Baulch et al. (2005) \\ 
$\ce{CH3OH}   +\ce{O}  \longrightarrow\ce{OH}     +\ce{CH2OH}        $ & $2.05\times10^{-11}$ & 0.00 &   2670.0 & Baulch et al. (2005) \\ 
$\ce{CH3OH}   +\ce{O}  \longrightarrow\ce{OH}     +\ce{CH3O}         $ & $2.05\times10^{-11}$ & 0.00 &   2670.0 & Baulch et al. (2005) \\ 
$\ce{CH3OH}   +\ce{OH} \longrightarrow\ce{CH2OH}  +\ce{H2O}          $ & $5.00\times10^{-13}$ & 1.92 &   -144.0 & Baulch et al. (2005) \\ 
$\ce{CH3OH}   +\ce{OH} \longrightarrow\ce{CH3O}   +\ce{H2O}          $ & $8.84\times10^{-14}$ & 1.92 &   -144.0 & Baulch et al. (2005) \\ 
$\ce{CH3OH}   +\ce{CH3}\longrightarrow\ce{CH4}    +\ce{CH2OH}        $ & $5.87\times10^{-15}$ & 3.45 &   4020.0 & Baulch et al. (2005) \\ 
$\ce{CH3OH}   +\ce{CH3}\longrightarrow\ce{CH4}    +\ce{CH3O}         $ & $1.17\times10^{-14}$ & 3.45 &   4020.0 & Baulch et al. (2005) \\ 
$\ce{CH2OH}   +\ce{H}  \longrightarrow\ce{H2CO}   +\ce{H2}           $ & $4.06\times10^{-11}$ & 0.00 &      0.0 & Baulch et al. (2005) \\ 
$\ce{CH2OH}   +\ce{H}  \longrightarrow\ce{CH3}    +\ce{OH}           $ & $1.74\times10^{-11}$ & 0.00 &      0.0 & Baulch et al. (2005) \\ 
$\ce{CH2OH}   +\ce{O}  \longrightarrow\ce{H2CO}   +\ce{OH}           $ & $7.01\times10^{-11}$ & 0.00 &      0.0 & Tsang (1987)\\ 
$\ce{CH2OH}   +\ce{OH} \longrightarrow\ce{H2CO}   +\ce{H2O}          $ & $4.00\times10^{-11}$ & 0.00 &      0.0 & Tsang (1987) \\ 
$\ce{CH2OH}   +\ce{HCO}\longrightarrow\ce{CH3OH}  +\ce{CO}           $ & $2.00\times10^{-10}$ & 0.00 &      0.0 & Tsang (1987) \\ 
$\ce{CH2OH}   +\ce{HCO}\longrightarrow\ce{H2CO}   +\ce{H2CO}         $ & $3.00\times10^{-10}$ & 0.00 &      0.0 & Tsang (1987) \\ 
$\ce{CH2OH}   +\ce{CH3}\longrightarrow\ce{H2CO}   +\ce{CH4}          $ & $4.00\times10^{-12}$ & 0.00 &      0.0 & Tsang (1987) \\ 
$\ce{CH2OH}   +\ce{O2} \longrightarrow\ce{H2CO}   +\ce{O2H}          $ & $9.24\times10^{-12}$ &-1.50 &      0.0 & Baulch et al. (2005) \\ 
$\ce{CH2OH}   +\ce{O2} \longrightarrow\ce{H2CO}   +\ce{O2H}          $ & $1.20\times10^{-10}$ & 0.00 &   1800.0 & Baulch et al. (2005) \\ 
$\ce{CH3O}    +\ce{H}  \longrightarrow\ce{H2CO}   +\ce{H2}           $ & $6.30\times10^{-11}$ & 0.00 &    300.0 & Baulch et al. (2005) \\ 
$\ce{CH3O}    +\ce{H}  \longrightarrow\ce{CH3}    +\ce{OH}           $ & $2.70\times10^{-11}$ & 0.00 &    300.0 & Baulch et al. (2005) \\ 
$\ce{CH3O}    +\ce{O}  \longrightarrow\ce{O2}     +\ce{CH3}          $ & $1.88\times10^{-11}$ & 0.00 &      0.0 & Baulch et al. (2005) \\ 
$\ce{CH3O}    +\ce{O}  \longrightarrow\ce{OH}     +\ce{H2CO}         $ & $6.25\times10^{-12}$ & 0.00 &      0.0 & Baulch et al. (2005) \\ 
$\ce{CH3O}    +\ce{OH} \longrightarrow\ce{H2CO}   +\ce{H2O}          $ & $3.01\times10^{-11}$ & 0.00 &      0.0 & Tsang \& Hampson (1986) \\ 
$\ce{CH3O}    +\ce{HCO}\longrightarrow\ce{CH3OH}  +\ce{CO}           $ & $1.50\times10^{-10}$ & 0.00 &      0.0 & Tsang \& Hampson (1986) \\ 
$\ce{CH3O}    +\ce{CH3}\longrightarrow\ce{H2CO}   +\ce{CH4}          $ & $4.00\times10^{-11}$ & 0.00 &      0.0 & Tsang \& Hampson (1986) \\ 
$\ce{CH3O}    +\ce{O2} \longrightarrow\ce{H2CO}   +\ce{O2H}          $ & $3.60\times10^{-14}$ & 0.00 &    880.0 & Baulch et al. (2005) \\ 
$\ce{CH3OCH3} +\ce{H}  \longrightarrow\ce{CH3OCH2}+\ce{H2}           $ & $1.47\times10^{-13}$ & 4.00 &    926.0 & Takahashi et al. (2007) \\ 
$\ce{CH3OCH3} +\ce{O}  \longrightarrow\ce{CH3OCH2}+\ce{OH}           $ & $8.30\times10^{-11}$ & 0.00 &   2300.0 & Herron (1988) \\ 
$\ce{CH3OCH3} +\ce{OH} \longrightarrow\ce{CH3OCH2}+\ce{H2O}          $ & $1.58\times10^{-12}$ & 1.73 &    176.0 & Baulch et al. (2005) \\ 
$\ce{CH3OCH3} +\ce{CH3}\longrightarrow\ce{CH3OCH2}+\ce{CH4}          $ & $4.35\times10^{-14}$ & 2.68 &   4190.0 & Baulch et al. (2005) \\ 
$\ce{CH3OCH2} +\ce{H}  \longrightarrow\ce{CH3CHO} +\ce{H2}           $ & $3.32\times10^{-11}$ & 0.00 &      0.0 & Edelb\"{u}ttel-Einhaus et al. (1992) \\ 
$\ce{CH3OCH2} +\ce{O}  \longrightarrow\ce{HCOOCH3}+\ce{H}            $ & $2.56\times10^{-10}$ & 0.15 &      0.0 & Song et al. (2005) \\ 
$\ce{HCOOCH3} +\ce{H}  \longrightarrow\ce{CH3O}   +\ce{CO}   +\ce{H2}$ & $2.81\times10^{-12}$ & 2.05 &   4471.0 & Good \& Francisco (2002) \\ 
$\ce{HCOOCH3} +\ce{O}  \longrightarrow\ce{CH3O}   +\ce{CO}   +\ce{OH}$ & $1.49\times10^{-11}$ & 0.00 &   2700.0 & Herron (1988) \\ 
$\ce{HCOOCH3} +\ce{OH} \longrightarrow\ce{CH3O}   +\ce{CO}   +\ce{H2}$ & $3.22\times10^{-13}$ & 2.00 &    734.0 & Good \& Francisco (2002) \\ 
$\ce{HCOOCH3} +\ce{CH3}\longrightarrow\ce{CH3O}   +\ce{CO}   +\ce{CH}$ & $4.70\times10^{-15}$ & 3.35 &   5800.0 & Good \& Francisco (2002) \\ 
$\ce{CH3COCH3}+\ce{H}  \longrightarrow\ce{CH3CO}  +\ce{CH2}  +\ce{H2}$ & $3.44\times10^{-12}$ & 2.00 &   2516.0 & Sato \& Hidake (2000) \\ 
$\ce{CH3COCH3}+\ce{O}  \longrightarrow\ce{CH3CO}  +\ce{CH2}  +\ce{OH}$ & $1.66\times10^{-11}$ & 0.00 &   3000.0 & Herron (1988) \\ 
$\ce{CH3COCH3}+\ce{OH} \longrightarrow\ce{CH3CO}  +\ce{CH2}  +\ce{H2}$ & $2.81\times10^{-12}$ & 0.00 &    760.0 & Atkinson et al. (1997) \\ 
$\ce{CH3COCH3}+\ce{CH3}\longrightarrow\ce{CH3CO}  +\ce{CH2}  +\ce{CH}$ & $5.76\times10^{-13}$ & 0.00 &   4870.0 & Arthur \& Newitt (1985) \\ 
$\ce{CH3CHO}  +\ce{H}  \longrightarrow\ce{CH3CO}  +\ce{H2}           $ & $2.55\times10^{-12}$ & 1.16 &   1210.0 & Baulch et al. (2005) \\ 
$\ce{CH3CHO}  +\ce{H}  \longrightarrow\ce{C2H5}   +\ce{O}            $ & $2.55\times10^{-12}$ & 1.16 &   1210.0 & Baulch et al. (2005) \\ 
$\ce{CH3CHO}  +\ce{O}  \longrightarrow\ce{CH3CO}  +\ce{OH}           $ & $9.70\times10^{-12}$ & 0.00 &    910.0 & Baulch et al. (2005) \\ 
$\ce{CH3CHO}  +\ce{O2} \longrightarrow\ce{O2H}    +\ce{CH3CO}        $ & $3.12\times10^{-13}$ & 2.50 &  18900.0 & Baulch et al. (2005) \\ 
$\ce{CH3CHO}  +\ce{OH} \longrightarrow\ce{CH3CO}  +\ce{H2O}          $ & $1.06\times10^{-12}$ & 1.35 &   -792.0 & Baulch et al. (2005) \\ 
$\ce{CH3CHO}  +\ce{O2H}\longrightarrow\ce{H2O2}   +\ce{CH3CO}        $ & $1.06\times10^{-13}$ & 2.50 &   5135.0 & Baulch et al. (2005) \\ 
$\ce{CH3CHO}  +\ce{CH3}\longrightarrow\ce{CH4}    +\ce{CH3CO}        $ & $1.40\times10^{-16}$ & 6.21 &    820.0 & Baulch et al. (2005) \\ 
$\ce{CH3CO}   +\ce{H}  \longrightarrow\ce{CH2CO}  +\ce{H2}           $ & $1.92\times10^{-11}$ & 0.00 &      0.0 & Warnatz (1984) \\ 
$\ce{CH3CO}   +\ce{H}  \longrightarrow\ce{HCO}    +\ce{CH3}          $ & $1.03\times10^{-11}$ & 0.00 &      0.0 & Warnatz (1984) \\ 
$\ce{CH3CO}   +\ce{O}  \longrightarrow\ce{CH2CO}  +\ce{OH}           $ & $8.75\times10^{-11}$ & 0.00 &      0.0 & Baulch et al. (2005) \\ 
$\ce{CH3CO}   +\ce{O}  \longrightarrow\ce{CO2}    +\ce{CH3}          $ & $2.63\times10^{-10}$ & 0.00 &      0.0 & Baulch et al. (2005) \\ 
$\ce{CH3CO}   +\ce{OH} \longrightarrow\ce{CH2CO}  +\ce{H2O}          $ & $2.01\times10^{-11}$ & 0.00 &      0.0 & Tsang \& Hampson (1986) \\ 
\hline
\end{tabular}
\begin{list}{}{}
\item[] {\bf Notes:} $k = \alpha(T/300)^{\beta}\exp{(-\gamma/T)}\,\mathrm{cm}^{3}\,\mathrm{s}^{-1}$ and M indicates a collisional partner.
\end{list}
}
\end{center}
\label{tableA1}
\end{table*}
\end{footnotesize}

\begin{footnotesize}
\begin{table*}
\caption{Reactions added to {\sc Rate}12 (continued).}
\begin{center}
{\small
\begin{tabular}{lcccc}
\noalign{\smallskip}
\hline\noalign{\smallskip}
              & $\alpha$ &         & $\gamma$      &  \\       
\multicolumn{1}{c}{Reaction}      &  (cm$^{3}\,$s$^{-1}$)              & $\beta$ & (K) & Reference \\ 
\noalign{\smallskip}
\hline\noalign{\smallskip}
$\ce{CH3CO}   +\ce{CH3}\longrightarrow\ce{CH2CO}  +\ce{CH4}          $ & $1.01\times10^{-11}$ & 0.00 &      0.0 & Hassinen et al. (1990) \\ 
$\ce{CH3CO}   +\ce{H2} \longrightarrow\ce{CH3CHO} +\ce{H}            $ & $2.20\times10^{-13}$ & 1.82 &   8860.0 & Tsang \& Hampson (1986) \\ 
$\ce{CH3CO}   +\ce{CH4}\longrightarrow\ce{CH3CHO} +\ce{CH3}          $ & $4.91\times10^{-14}$ & 2.88 &  10800.0 & Tsang \& Hampson (1986) \\ 
$\ce{CH3CO}   +\ce{CH2}\longrightarrow\ce{CH2CO}  +\ce{CH3}          $ & $3.01\times10^{-11}$ & 0.00 &      0.0 & Tsang \& Hampson (1986) \\ 
$\ce{CH2CO}   +\ce{O}  \longrightarrow\ce{H2CO}   +\ce{CO}           $ & $6.00\times10^{-13}$ & 0.00 &    680.0 & Baulch et al. (2005) \\ 
$\ce{CH2CO}   +\ce{O}  \longrightarrow\ce{HCO}    +\ce{CO}   +\ce{H} $ & $3.00\times10^{-13}$ & 0.00 &    680.0 & Baulch et al. (2005) \\ 
$\ce{CH2CO}   +\ce{O}  \longrightarrow\ce{HCO}    +\ce{HCO}          $ & $3.00\times10^{-13}$ & 0.00 &    680.0 & Baulch et al. (2005) \\ 
$\ce{CH2CO}   +\ce{O}  \longrightarrow\ce{CO2}    +\ce{CH2}          $ & $1.80\times10^{-12}$ & 0.00 &    680.0 & Baulch et al. (2005) \\ 
$\ce{CH2CO}   +\ce{OH} \longrightarrow\ce{CH3}    +\ce{CO2}          $ & $1.04\times10^{-12}$ & 0.00 &   -510.0 & Baulch et al. (2005) \\ 
$\ce{CH2CO}   +\ce{OH} \longrightarrow\ce{CH2OH}  +\ce{CO}           $ & $1.68\times10^{-12}$ & 0.00 &   -510.0 & Baulch et al. (2005) \\ 
$\ce{CH2CO}   +\ce{OH} \longrightarrow\ce{H2CO}   +\ce{HCO}          $ & $5.60\times10^{-14}$ & 0.00 &   -510.0 & Baulch et al. (2005) \\ 
$\ce{CH2CO}   +\ce{OH} \longrightarrow\ce{HC2O}   +\ce{H2O}          $ & $2.80\times10^{-14}$ & 0.00 &   -510.0 & Baulch et al. (2005) \\ 
$\ce{HC2O}    +\ce{H}  \longrightarrow\ce{CO}     +\ce{CH2}          $ & $4.98\times10^{-11}$ & 0.00 &      0.0 & Warnatz (1984) \\ 
$\ce{HC2O}    +\ce{O}  \longrightarrow\ce{CO2}    +\ce{CH}           $ & $1.99\times10^{-12}$ & 0.00 &      0.0 & Warnatz (1984) \\ 
$\ce{H2CO}    +\ce{M}  \longrightarrow\ce{HCO}    +\ce{H}    +\ce{M} $ & $2.45\times10^{-08}$ & 0.00 &  38050.0 & Baulch et al. (2005) \\ 
$\ce{H2CO}    +\ce{M}  \longrightarrow\ce{CO}     +\ce{H2}   +\ce{M} $ & $1.42\times10^{-08}$ & 0.00 &  32100.0 & Baulch et al. (2005) \\ 
$\ce{HNCO}    +\ce{M}  \longrightarrow\ce{CO}     +\ce{NH}   +\ce{M} $ & $2.24\times10^{-03}$ &-3.10 &  51300.0 & Tsang (1992) \\ 
$\ce{CH2OH}   +\ce{M}  \longrightarrow\ce{H2CO}   +\ce{H}    +\ce{M} $ & $1.38\times10^{-04}$ &-2.50 &  17200.0 & Tsang (1987) \\ 
$\ce{CH3O}    +\ce{M}  \longrightarrow\ce{H2CO}   +\ce{H}    +\ce{M} $ & $2.59\times10^{-10}$ & 0.00 &   6790.0 & Baulch et al. (1994) \\ 
$\ce{CH3OH}   +\ce{M}  \longrightarrow\ce{CH3}    +\ce{OH}   +\ce{M} $ & $3.38\times10^{-07}$ & 0.00 &  33100.0 & Baulch et al. (1994) \\ 
$\ce{HC2O}    +\ce{M}  \longrightarrow\ce{CO}     +\ce{CH}   +\ce{M} $ & $3.52\times10^{-08}$ & 0.00 &  29600.0 & Frank et al. (1988) \\ 
$\ce{CH2CO}   +\ce{M}  \longrightarrow\ce{CO}     +\ce{CH2}  +\ce{M} $ & $1.25\times10^{-08}$ & 0.00 &  29000.0 & Frank et al. (1988) \\ 
$\ce{CH3CO}   +\ce{M}  \longrightarrow\ce{CO}     +\ce{CH3}  +\ce{M} $ & $1.38\times10^{-08}$ & 0.00 &   7080.0 & Baulch et al. (2005) \\ 
$\ce{HCOOH}   +\ce{M}  \longrightarrow\ce{CO}     +\ce{H2O}  +\ce{M} $ & $2.78\times10^{-09}$ & 0.00 &  21800.0 & NIST Fit \\ 
$\ce{HCOOH}   +\ce{M}  \longrightarrow\ce{CO2}    +\ce{H2}   +\ce{M} $ & $1.12\times10^{-08}$ & 0.00 &  25800.0 & NIST Fit \\ 
$\ce{CH3OCH3} +\ce{M}  \longrightarrow\ce{CH3O}   +\ce{CH3}  +\ce{M} $ & $4.33\times10^{-08}$ & 0.00 &  21537.0 & Cook et al. (2009) \\ 
$\ce{CH3CHO}  +\ce{M}  \longrightarrow\ce{CH3}    +\ce{HCO}  +\ce{M} $ & $5.63\times10^{-04}$ & 0.00 &  37100.0 & Trenwith (1963) \\ 
$\ce{CH3OCH2} +\ce{M}  \longrightarrow\ce{H2CO}   +\ce{CH3}          $ & $1.60\times10^{-07}$ & 0.00 &   9100.0 & Loucks \& Laidler (1967) \\ 
$\ce{HCOOCH3} +\ce{M}  \longrightarrow\ce{CH3O}   +\ce{HCO}  +\ce{M} $ & $1.00\times10^{-10}$ & 0.00 &  43060.0 & Willacy et al. (1998) \\   
\hline
\end{tabular}
\begin{list}{}{}
\item[] {\bf Notes:} $k = \alpha(T/300)^{\beta}\exp{(-\gamma/T)}\,\mathrm{cm}^{3}\,\mathrm{s}^{-1}$ and M indicates a collisional partner.
\end{list}
}
\end{center}
\label{tableA1}
\end{table*}
\end{footnotesize}


\end{appendix}

\end{document}